%% file: arxiv_manu.tex
\documentclass[conference]{IEEEtran}
\IEEEoverridecommandlockouts
\usepackage{cite}
\usepackage{amsmath,amssymb,amsfonts}
\usepackage{algorithmic}
\usepackage{graphicx}
\usepackage{epstopdf}
\usepackage{multirow}
\usepackage{subcaption}
\usepackage{gensymb}
\usepackage{textcomp}
\usepackage{xcolor}
\usepackage{tikz,tikz-qtree}
\usetikzlibrary{positioning,fit,calc}
\usetikzlibrary{fit}
\tikzset{block/.style={draw,thick,text width=3cm,minimum height=1cm,align=center},
         line/.style={-latex}
}
\tikzset{block1/.style={draw,thick,text width=2cm,minimum height=1cm,align=center},
         line/.style={-latex}
}
\tikzset{block2/.style={draw,thick,text width=2cm,minimum height=2cm,align=center},
         line/.style={-latex}
}
\def\BibTeX{{\rm B\kern-.05em{\sc i\kern-.025em b}\kern-.08em
    T\kern-.1667em\lower.7ex\hbox{E}\kern-.125emX}}
\begin{document}

\title{Identification/Segmentation of Indian Regional Languages with Singular Value Decomposition based Feature Embedding}

\author{\IEEEauthorblockN{ Anirban Bhowmick}
\IEEEauthorblockA{
\textit{VIT Bhopal University}\\
Sehore, India \\
anirban.bhowmick@vitbhopal.ac.in}
\and
\IEEEauthorblockN{Astik Biswas}
\IEEEauthorblockA{
\textit{Stellenbosch University}\\
South Africa \\
abiswas@sun.ac.za}
}
\maketitle

\begin{abstract}
language identification (LID) is identifing a language in a given spoken utterance. 
Language segmentation is equally inportant as language identification where language boundaries
can be spotted in a multi language utterance. In this paper, we have experimented with two schemes for language identification in Indian regional language context as very few works has been done. Singular value based feature embedding is used for both of the schemes. In first scheme, the singular value decomposition (SVD) is applied to the n-gram utterance matrix and in the second scheme, SVD is applied on the difference supervector matrix space. We have observed that in
both the schemes, 55-65\% singular value energy is sufficient to capture the language context. In n-gram based feature representation, we have seen that different skipgram models captures different language context. We have observed that for short test duration, supervector based feature representation is
better but with a longer duration test signal, n-gram based feature performed better. We have also extended our work to explore language based segmentation where we have seen that segmentation accuracy of four language group with ten language training model, scheme-1 has performed well but with same four language training model, scheme-2 outperformed scheme-1.

\end{abstract}


\section{Introduction}
Automatic language identification (LID) is the process of identifying the language of a given spoken utterance \cite{muthusamy1994automatic}. LID system has increasing importance among speech processing applications. It can be used in multilingual translation systems or emergency call routing, pre-selecting suitable speech recognition system \cite{ambikairajah2011language,lopez2014automatic}. The LID task can be carried out with different levels of explicit language description (phones, prosody) or implicit LID system, which doesn't require any label corpora to perform the identification \cite{gelly2017spoken}. \par
Phonotactic systems are the earliest Language identification system where the use of phone-based acoustic likelihoods was proposed \cite{lamel1993identifying,lamel1994language}.
Parallel phone recognizers with language model (PPRLM) \cite{zissman1994automatic,zissman1996comparison} is the example of the explicit LID system. To provide state-of-the-art results  different methods based on phone decoding with phonotactic methods have been explored in recent years \cite{benzeghiba2010improved,benzeghiba2012phonotactic,benzeghiba2012fusing} . Implicit LID systems do not require any explicit language description, GMM tokenization is an example of an implicit LID system \cite{torres2002approaches}.\par
Acoustic modelling based techniques have been always meaningful in language identification. Implementation of GMM-UBM based speaker verification (SV) system has been used in language identification \cite{you2010gmm,kumar2015significance}. Shifted delta cepstral (SDC) feature which capture language dynamics, was introduced in language identification \cite{torres2002approaches,wang2013shifted}. To reduce the training and testing data mismatch joint factor analysis (JFA) was explored. JFA tries to remove intersession variability by modelling that subspace. But, the problem with JFA is that the channel factor also contains speaker information, for that total variability (TV) space was proposed to solve this problem. I-vector formulation from the TV space is the state-of-the-art LID system. Essentially the idea is borrowed from the popular SV technique.
Recently, LID work has been explored in deep neural network based work, where long short time memory (LSTM) based recurrent neural network (RNN) has been used to capture and model the fine temporal variation in a language. Although, much has been done in the neural network based framework but acoustic model based technique is still a success with limited amount of data. \par
In this work we have explored two schemes for language identification. We have performed our experiments with Indian regional languages as very little work has been done in this context. We have used Gaussian mixture model (GMM)- universal background model supervector and n-gram based feature representation with singular value decomposition based feature embedding to capture the language variability among the other variabilities (speaker, channel). Eigenvoice which is inspired from eigenfaces has applied principle component analysis (PCA) on the class dependent hidden Markov model (HMM)-GMM supervectors. We have taken the same approach but instead of making class dependent model we have built class-independent models. For speaker and channel normalization, we have used ceptral mean normalization (CMN) with an estimated sliding window. Instead of using PCA based feature embedding, we have used SVD. In both of the schemes, SVD is applied. For first scheme, the SVD is applied to the n-gram utterance matrix and for second scheme on the difference supervector matrix space  The top singular vectors are selected from the singular value distribution, to retain the maximum language information. It has been observed that 55-65\% singular value energy is sufficient to capture the language context. With skipgram models incorporated with n-gram based feature representation, it has been observed that different language contexts can be captured. For short test duration signals, supervector based feature representation performs well whereas, for the longer test duration signals, n-gram based feature representation performs better. This language identification has also been extended to language speech segmentation.
\section{Language Identification/Segmentation Schemes}
We have investigated two schemes for language identification/segmentation: n-gram based technique and difference matrix based technique. The block diagram is showing the training phase for both the schemes. In the Fig.\ref{fig:train_dia}, we can see the initial block diagrams are same for the both the schemes, the difference is in feature representation, one scheme is based on n-gram matrix based feature representation (Method 1) and the other scheme is based on difference matrix based feature representation (Method 2).  
\input{train_diagram}
\subsection{Feature extraction and universal background model(UBM)}
Raw speech signal is parametrized using 39 dimensional MFCC features (13+13$\Delta$ +13$\Delta\Delta$), with frame size of 25 ms and a  overlap of 10 ms. Cepstral mean normalization (CMN) is performed to normalize the feature vectors. The optimal value of sliding window in CMN is calculated by estimating the Fisher's linear discriminant.\par Fisher's linear discriminant is based on the analysis of two scatter matrices: within-class scatter matrix and between-class scatter matrix. 
Given a set of samples $s_1,\dots,s_n$ and their class labels $y_1,\dots,y_n$, The within-class scatter matrix is defined as:
\begin{equation}
S_w=\sum_{i=1}^{n} (x_i-\mu_{y_{i}})(x_i-\mu_{y_{i}})^{T}
\end{equation}
where, $\mu_k$ is the sample mean for $k^{th}$ class. The between-class scatter matrix is defined as:
\begin{equation}
S_b=\sum_{k=1}^{m}n_k (\mu_k-\mu)(\mu_k-\mu)^{T}
\end{equation}
Here, $m$ is the number of classes, $\mu$ is the overall sample mean, and $n_k$ is the number of samples in the k-th class. The ratio of the between-class scattering ($S_b$) to the within-class scattering ($S_w$) is used to estimate Fisher measure. This has the effect of reducing channel variability and some of the speaker variability.
\begin{figure}[h]
    \centering
    \includegraphics[width=0.8\linewidth]{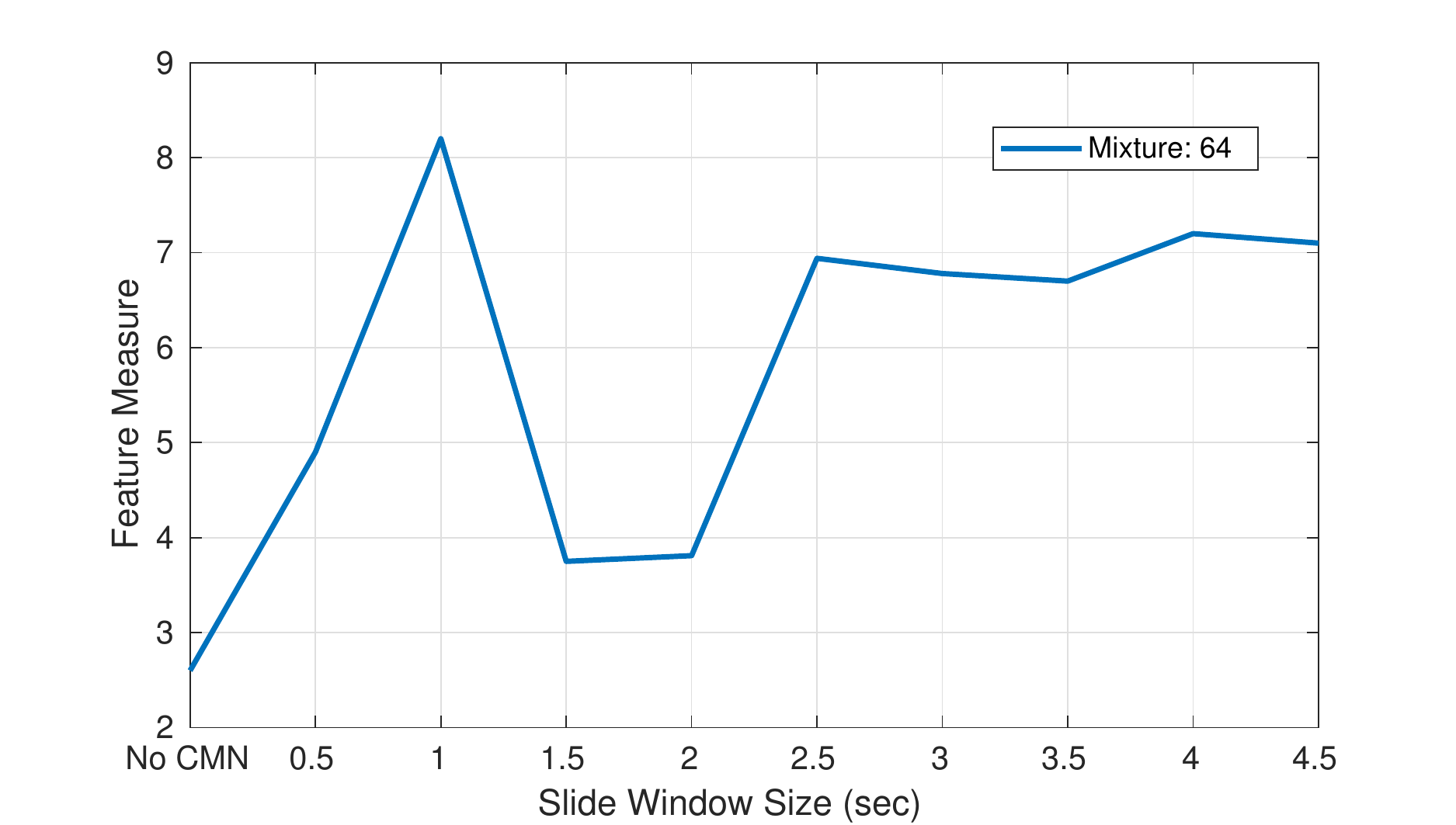}
    \caption{Sliding window length estimation for CMN  \label{fig:CMN}}
\end{figure} 
In Fig \ref{fig:CMN}, we can see that the 1 sec window duration is having higher Fisher value.
The language and speaker independent data is modelled using a mixture model such as Gaussian mixture model (GMM). This model is referred to as universal background model (UBM). We have used same number of speakers as well as same duration of data for each language. To build GMM-UBM, total 960 minutes of data has been used (10 language, 8 speakers/language, 4 session/speaker and 3 minutes/session).\par 
A UBM in the LID system is a GMM representing the characteristic of all different languages. Instead of employing the maximum-likelihood (ML) training, each language model can be created by performing Bayesian adaptation from the UBM using language specific training speech. The UBM-GMM system can be represented as,
\begin{equation}
\Lambda^{M}_{UBM}=\{w^{m}_{UBM},\mu^{m}_{UBM},\Sigma^{m}_{UBM}\}^{M}_{m=1}
\end{equation}
where, $M$ is the number of mixtures. 
In this GMM-UBM system, the hypothesized language model is derived by adapting the language data. In the normal ML estimation, the training of the language model is not dependent on UBM. But in the adaptation approach, language model parameters are derived by updating the models and models are trained by maximum a posteriori (MAP) estimation.
\subsection{ MAP adaptation}
MAP algorithm is a two step process, in the initial step, we compute training data's sufficient statistics for each of mixtures in the prior model and in the second step, using a data-dependent mixing coefficient, this new statistics are then mixed with old sufficient statistics from the prior mixture parameters. In final parameter estimation, data dependent mixing coefficient is used so that mixtures with more new data depends more on the new sufficient statistics and mixtures with less new data depends more on the old sufficient statistics.\par
The specifics of the adaptation are as follows: Given a UBM ( $\Lambda^{M}_{UBM}$) and utterance feature vectors $X$=$\{{x_{1},..,x_{T}}\}$, we determine the probabilistic alignment of the training vectors into the UBM mixture components to get utterance wise adapted models. Utterance-wise MAP adapted models are represented as:
\begin{equation*}
 \lambda_{lang(1:L)\_spk(1:N)\_sess(1:S)}=
 \end{equation*}
 \begin{equation}  
 \{\lambda_{lang(1)\_spk(1)\_sess(1)}, .... ,\lambda_{lang(L)\_spk(N)\_sess(S)}\}
\end{equation}
where, $L$ is the total number of language, $N$ is the number of speaker for that language and $S$ is the number of session per speaker.
\subsection{Scheme 1: Decoding Sequence \& N-gram Modelling}
After the MAP adaptation, Symbol sequence is decoded given the utterance feature vector of $t^{th}$ frame ($x_t$) and the adapted model  ($\lambda_{lang(l)\_spk(n)\_sess(s)}$) from the same feature vector. 
\begin{equation}
SS_{t}=\operatorname*{arg\,max}_{m\in1:M} p(x_t|\lambda_{lang(l)\_spk(n)\_sess(s)})
\end{equation}
where $l\in(1:L),n\in(1:N)$ and $s\in(1:S)$.
After decoding the Gaussian sequences, we make utterance wise n-gram model. The sparse bigram model or skipgram models can be estimated as follows:
\begin{equation}
B_{utt}(i,j)=\frac{Count(SS_{t-K}=i,SS_{t}=j)}{Count(SS_{t-K}=i)}
\end{equation}
where, $K$ is the skip parameter and $K\in(1,...,P)$. $P$ is any integer value. When $K=1$, it is bigram model. Different skipgram models capture different temporal context in a language. Temporal context can vary between languages.\par     
We, transform the utterance specific skipgram matrix ($B_{utt}$) of $M\times M$ to a vector ($Bv_{utt}$) of [$MM\times 1$], we stack all the utterance specific n-gram vectors to form a utterance n-gram matrix, as:
\begin{equation}
\mathcal{A}=[Bv_1,....,Bv_{N_{tot}}]
\end{equation}
where $utt\in 1:N_{tot}$, $N_{tot}$ is the total number of utterances. 

\subsection{Scheme 2: Difference matrix}
Utterance-specific supervector, $m_u$ is obtained by concatenating the means of corresponding adapted GMMs. Supervector matrix are formed by stacking up these utterance-wise supervectors  $Y$, as\cite{jain2018svd}:
\begin{equation}
m_u=[\mu_1, \mu_2, ...... ,\mu_M ]^{T}_{Md\times1}
\end{equation}

\begin{equation}
Y=[m_1, m_2, ...... ,m_{Nt} ]^{T}_{{N_{tot}}\times{Md}}
\end{equation}

where, $N_{tot}$= total number of training data utterances, $d$ = feature vectors, and  $\mu_m$ = mean vector of the m-th mixture component $(1\leq m \leq M)$. The UBM mean ($U_m$) is subtracted from every supervector to center them  with respect to the UBM as shown below \cite{jain2018svd}:
\begin{equation}
\Delta Y=[m_1, m_2, ...... ,m_{N_{tot}} ]^{T} - [U_m, U_m, ...... ,U_m ]^{T}
\end{equation}
where, $\Delta Y$ is the difference matrix.
\subsection{Feature Embedding (SVD)}
\begin{figure}[h]
    \centering
    \includegraphics[width=0.5\linewidth]{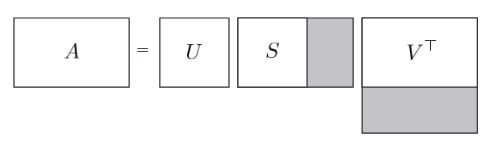}
    \caption{Singular value decomposition of a fat matrix  \label{fig:svd}}
\end{figure}

To reduce the dimensionality of the utterance n-gram matrix ($\mathcal{A}$) and difference matrix ($\Delta Y$), singular value decomposition (SVD) is performed, as:
\begin{equation}
\mathcal{A}=U_{1}S_{1}V_{1}^{T}
\end{equation}
\begin{equation}
{\Delta Y}=U_{2}S_{2}V_{2}^{T}
\end{equation}
where, $S_{1}$ and $S_{2}$ are the diagonal singular value matrix, $U_1$,$U_2$ and $V_1$,$V_2$ are the matrices consisting of left and right singular vectors. The dimension of the singular value is decided as:
\begin{equation}
\mathcal{R}_{1}=min \{ N_{tot},M\times M \}
\end{equation}
\begin{equation}
\mathcal{R}_{2}=min \{ N_{tot},M\times d \}
\end{equation}
As the dimension of $M\times M$ for utterance n-gram matrix and $M\times d$ for difference matrix are much higher than $N_{tot}$, we will get a fat matrix, and its decomposition will be similar to the Fig. \ref{fig:svd}. 
Singular values matrix $S_{1}$ and $S_{2}$ has dimension $N_{tot}\times N_{tot}$. If we plot the singular value distribution we can see that first few singular values capture most of the variation in the data, hence only first few singular values are enough to represent the language space.
\begin{figure}[h]
	\centering
	\begin{minipage}[b]{0.48\linewidth}
		\includegraphics[width=1\linewidth]{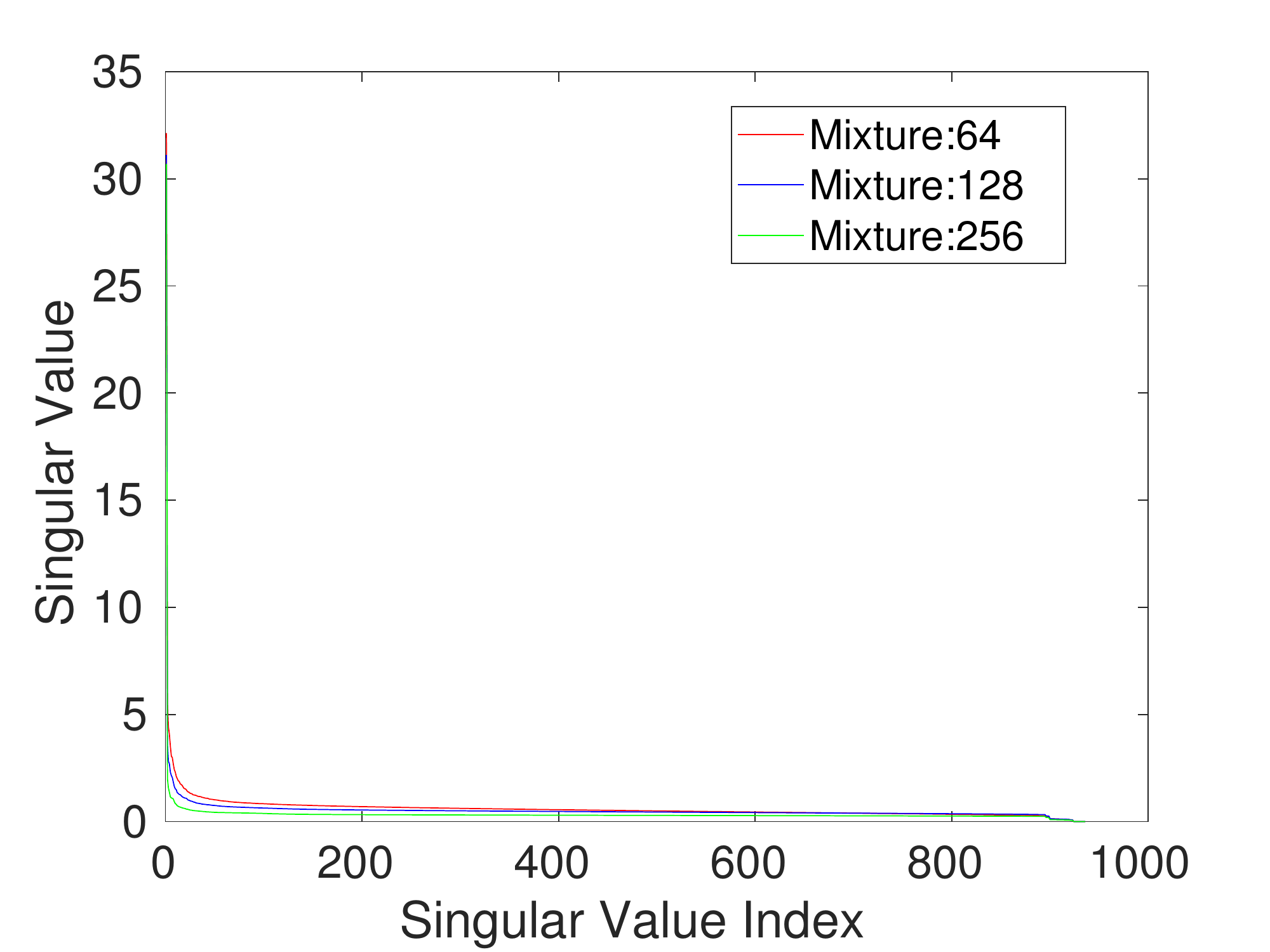}
		\centerline{(a)}
	\end{minipage}
		\begin{minipage}[b]{0.48\linewidth}
		\includegraphics[width=1\linewidth]{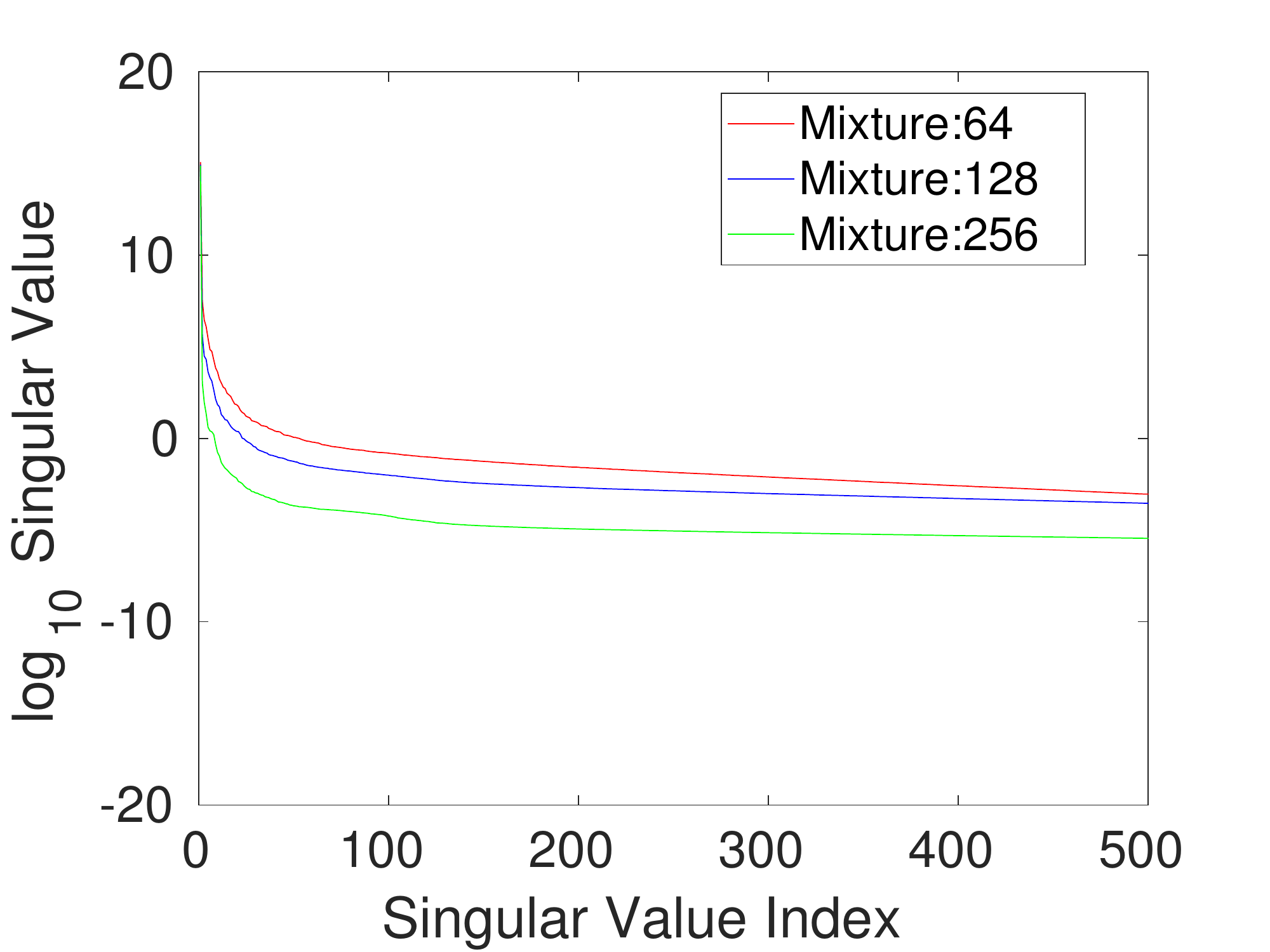}
		\centerline{(b)}		
	\end{minipage}
	\begin{minipage}[b]{0.48\linewidth}
		\includegraphics[width=1\linewidth]{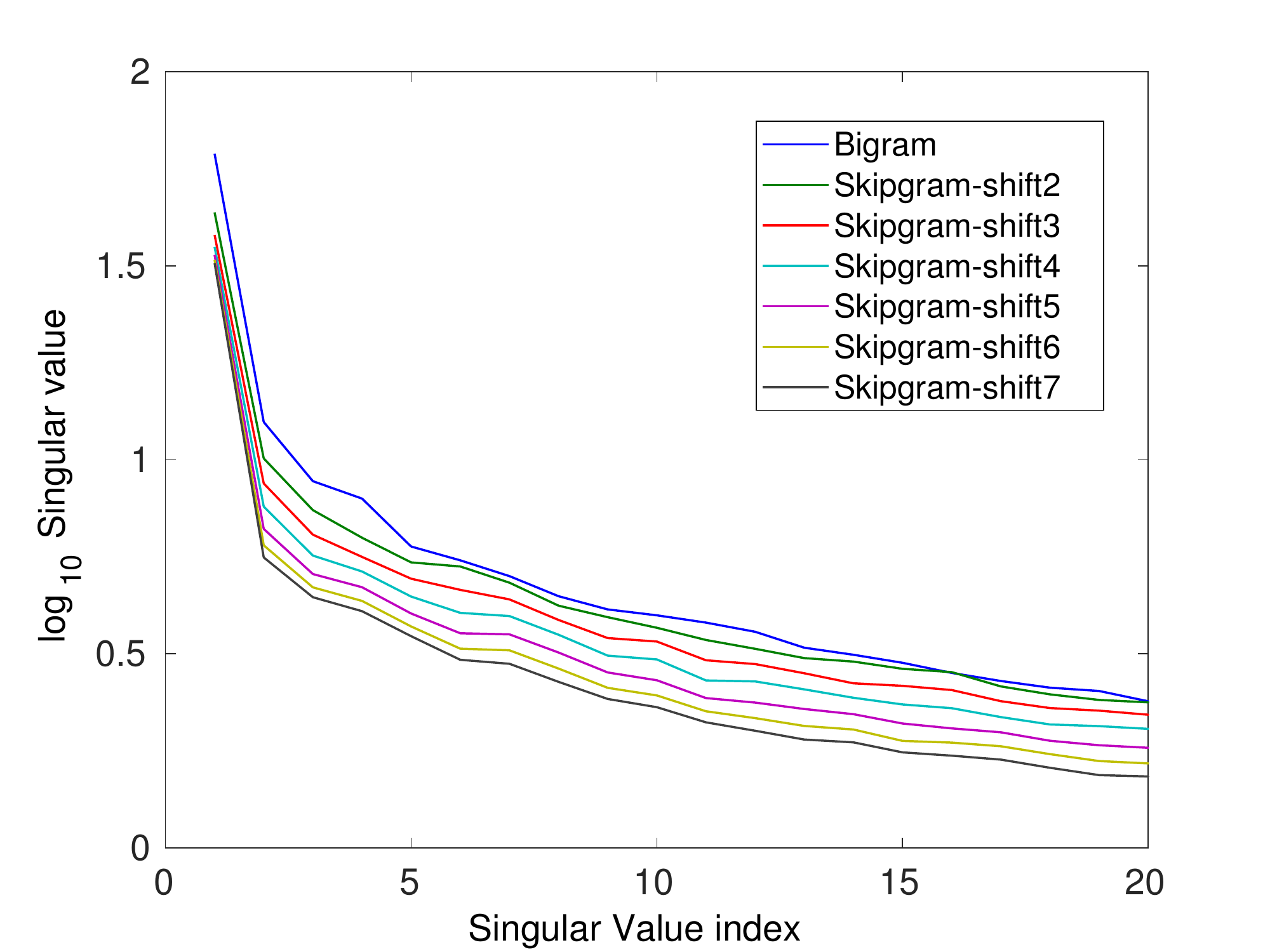}
		\centerline{(c)}		
	\end{minipage}
		\begin{minipage}[b]{0.48\linewidth}
		\includegraphics[width=1\linewidth]{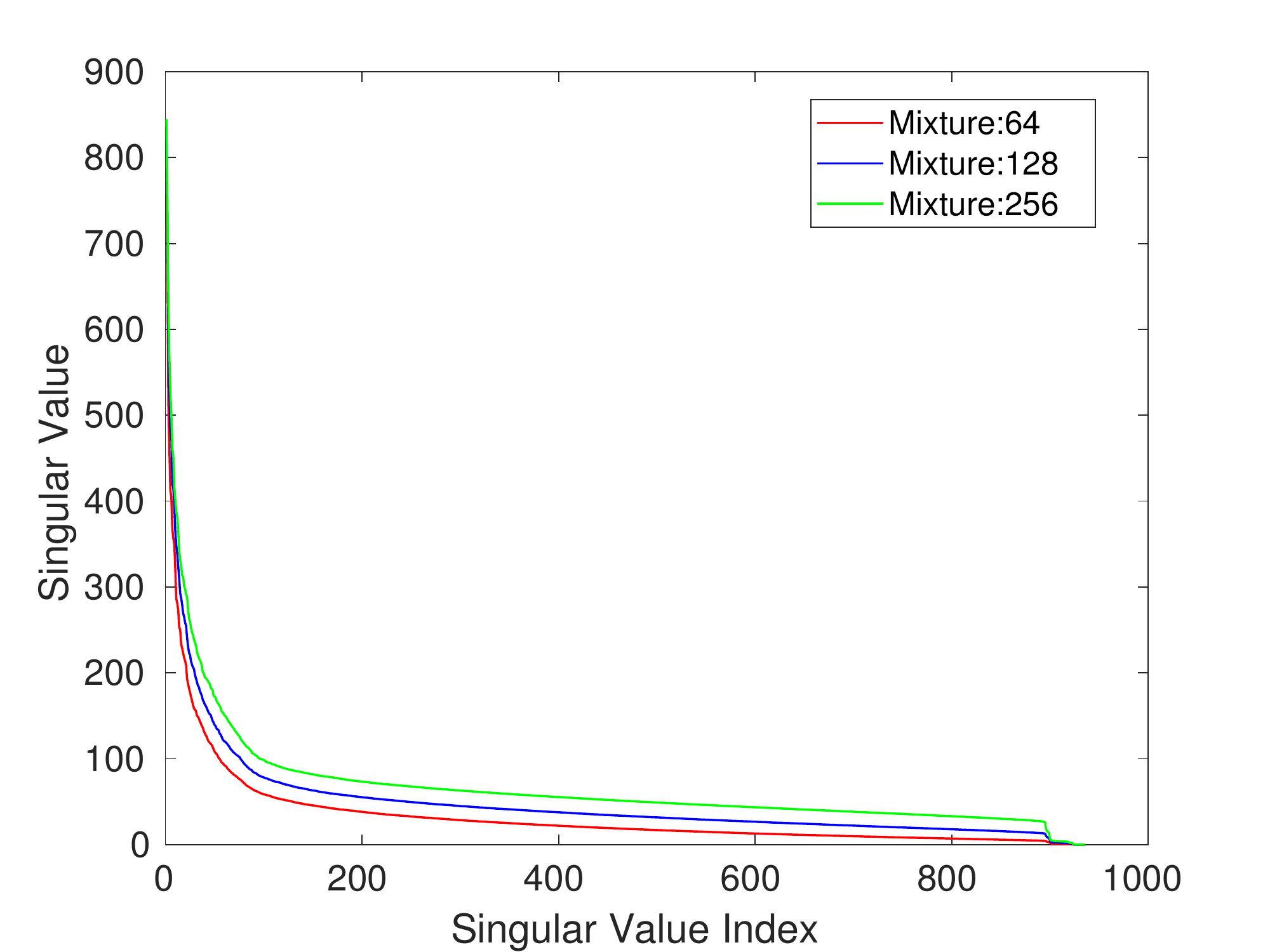}
		\centerline{(d)}		
	\end{minipage}
	\caption{(a) Singular value ($S_{1}$) distribution of n-gram utterance matrix  (b) Singular value distribution of utterance matrix at different value of M
(first 500)(skipgram-shift7) (c) Singular value distribution of different n-gram utterance matrix at $M$ = 64 (first 20 singular values) (d) Singular value ($S_{2}$) distribution of difference matrix}
	\label{fig:sing_dist}
\end{figure}
In Fig. \ref{fig:sing_dist}(a), we can see that the most of the energy is preserved in the initial singular values. The singular value distribution for different n-gram utterance matrix is almost similar. In Fig. \ref{fig:sing_dist}(b) and (c), change in singular value distribution with respect to change in n-gram models and with mixture components are shown. We have estimated that to preserve same amount of energy, the requirement of number of singular values increases with increased shift parameter of n-gram model and as well as with increase in mixture components. Similarly, for  \ref{fig:sing_dist}(d), we have found that to preserve same amount of energy we need more singular value dimensions. However, for singular value distribution $S_{2}$, energy/singular value is greater than $S_{1}$.\par
As, the initial singular values capture most of the variation in the data, hence, only first $\mathcal{L}_{1}$ out of $\mathcal{R}_{1}$ and $\mathcal{L}_{2}$ out of $\mathcal{R}_{2}$ values are enough to represent the language space. where, $\mathcal{L}_1 <  N_{tot} << MM$
and $\mathcal{L}_2 <  N_{tot} << Md$.
The singular values contain all the data variations, like language variation, speaker variation, channel variation etc. Hence, we have not chosen $\mathcal{L}_{1}$ and $\mathcal{L}_{2}$ based on the thumb rule. We have seen that adding more singular values actually deteriorate the identification accuracy. Hence, we have only retains 55-65\% of the energy in $S_{1}$ and $S_{2}$.\par

Now, the n-gram utterance matrix $\mathcal{A}$ of the training data is projected into lower dimensional space ($\mathcal{L}_{1}$) by projecting it along ${S_1}_{\mathcal{L}_{1}}$ and ${V_1}_{\mathcal{L}_{1}}$, as:
\begin{equation}
\mathcal{A}_{\mathcal{L}_{1}}=\mathcal{A}{V_1}_{\mathcal{L}_{1}}{S_1}_{\mathcal{L}_{1}}^{-1}
\end{equation}
The training utterances are projected to a lower dimensional space.  ${\mathcal{A}_{\mathcal{L}_{1}}}$ is of dimension $N_{tot}\times{\mathcal{L}_{1}}$.

Similarly, the difference matrix $\Delta Y$ of the training data is projected into lower dimensional space ($\mathcal{L}_{2}$) by projecting it along ${S_2}{\mathcal{L}_{2}}$ and ${V_2}_{\mathcal{L}_{2}}$, as:
\begin{equation}
\Delta Y_{\mathcal{L}_{2}}=\Delta Y{V_2}_{\mathcal{L}_{2}}{S_2}_{\mathcal{L}_{2}}^{-1}
\end{equation}
\begin{figure}[h]
	\centering
	\begin{minipage}[b]{0.48\linewidth}
		\includegraphics[width=1\linewidth]{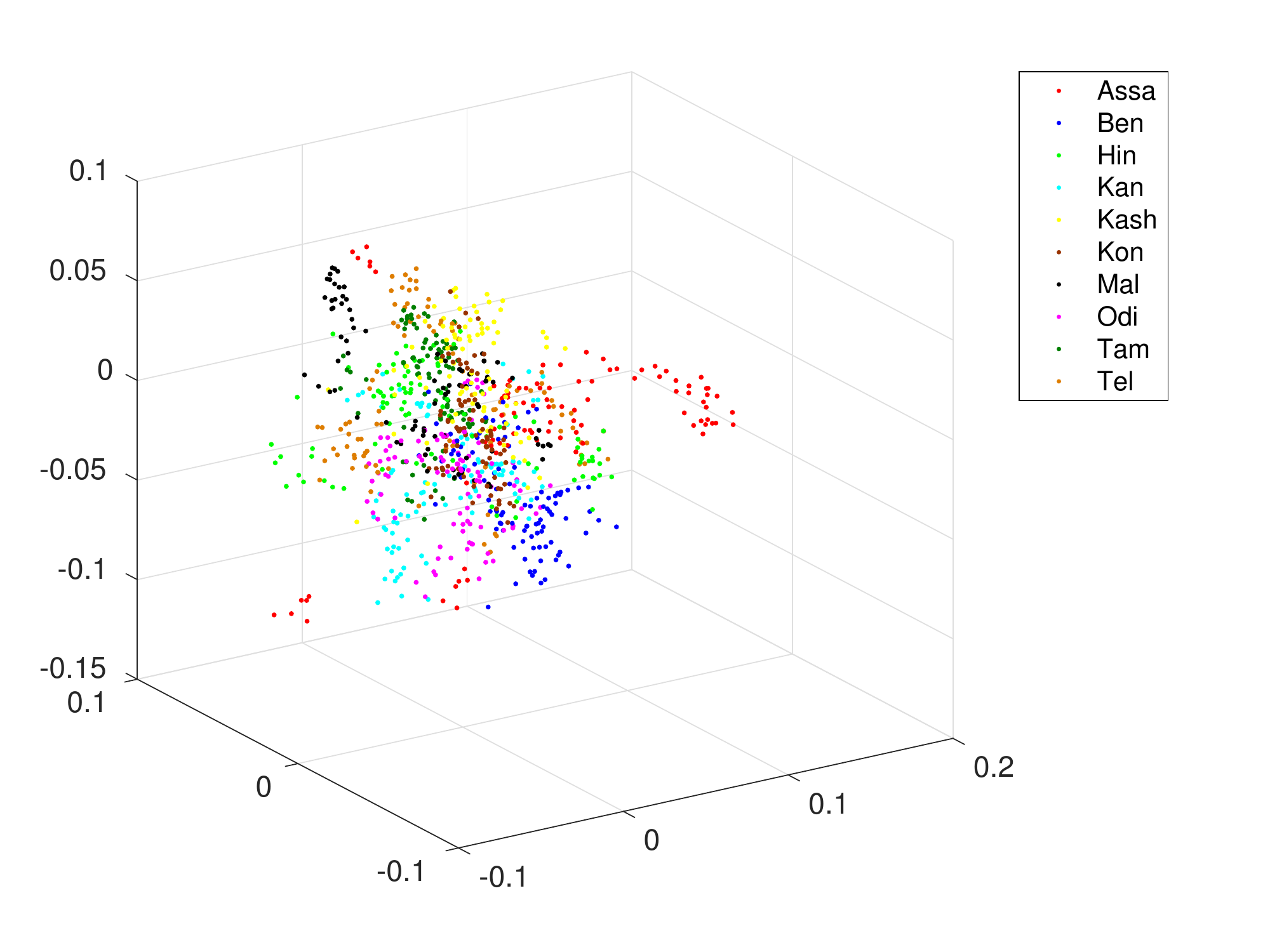}
		\centerline{(a)}
	\end{minipage}
	\begin{minipage}[b]{0.48\linewidth}
		\includegraphics[width=1\linewidth]{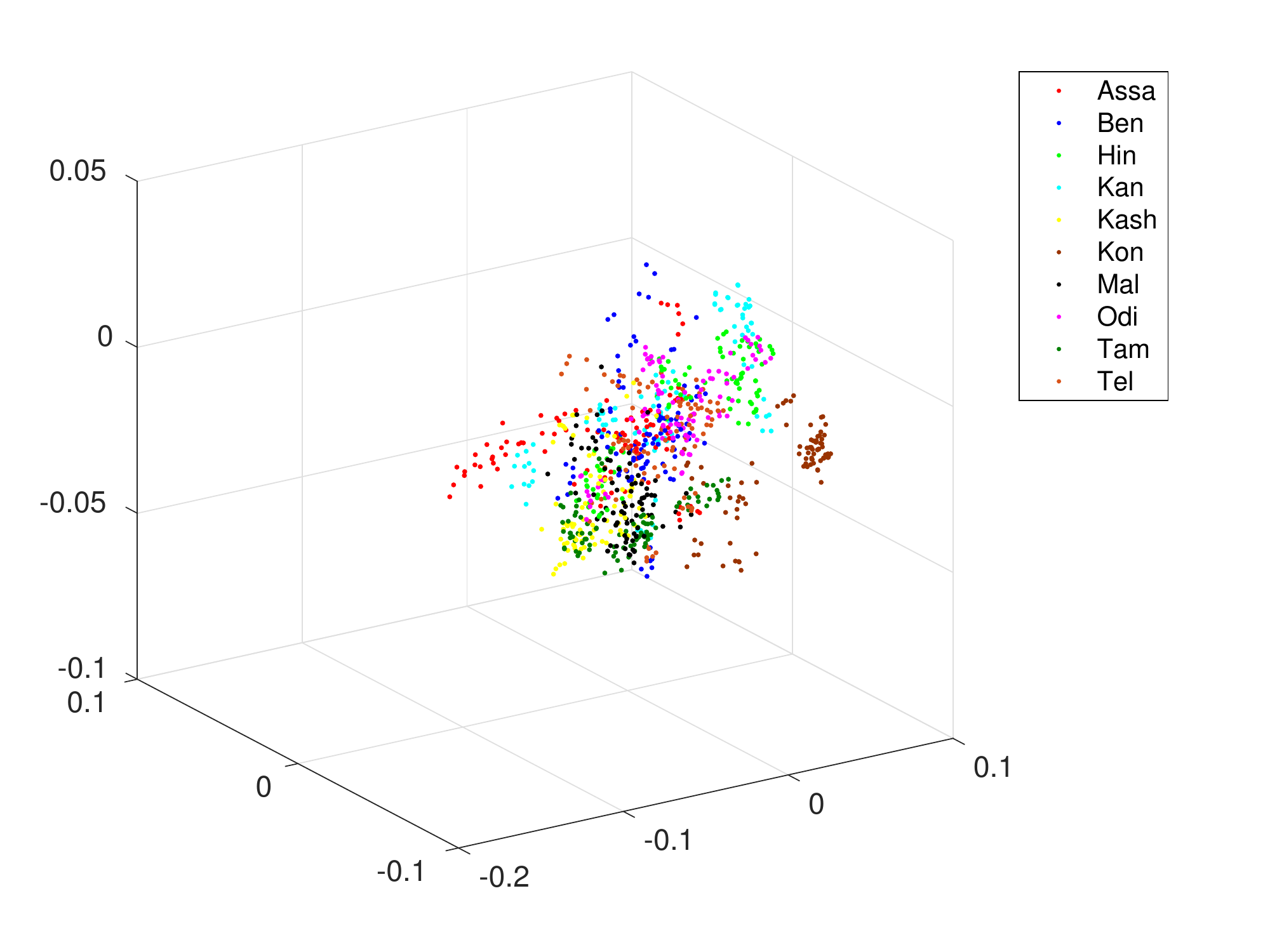}
		\centerline{(b)}		
	\end{minipage}
\caption{3D feature embedding of 10 language features with (a) Scheme-1 (b) Scheme-2}\label{fig:lang_fea}
\end{figure}

3 dimensional projection of the $\mathcal{A}_{\mathcal{L}_{1}}$ is shown in Fig. \ref{fig:lang_fea}(a), where we can see clustering of 10 languages.
Some languages shown better clustering than others. Assamese and Bengali clusters are nearby due to their similarity. Tamil and Malayalam also clustered at nearby place. Bengali and Malayalam language have larger inter cluster distance than Bengali and Odia or Bengali and Assamese.  
In Fig. \ref{fig:lang_fea}(b), we can see the 3 dimensional feature embedding of $\Delta Y_{\mathcal{L}_{2}}$. We can find the prominent clustering of Konkani language features. Bengali and Odia language features are fused in the same region. Language specific clustering will increase with increased projected dimension. 
\subsection{Support Vector Machine Modelling (SVM)}
After the projection of the matrix in a lower dimensional space, they are classified in a supervised manner using SVM. The number of class depends upon the number of languages to be classified. ``1 vs rest" approach is taken to discriminate all the languages. In this approach, $k$ SVM models are constructed where the class numbers are represented by $k$. 
all the examples in the $m^{th}$ class are used to train $m^{th}$ SVM with positive labels, and all other examples are used with negative labels. Thus given T training data $(x_1,y_1)$,...,$(x_T,y_T)$, where $x_i\in R^{N}$, $i=1,...,T$ and $y_i \in \{1,....k\}$ is the class of $x_i$. 
The formulation of SVM's optimization problem with slack variable $\xi$ is given below:
Find $w$,$b$ and $\xi\geq0$ such that 
\begin{equation}
\Phi=w^Tw + C\sum_{i=1}^{T}\xi^m_i
\end{equation}\label{eq.1}
Eq. \ref{eq.1} is minimized.
For $m^{th}$ class the problem:
\begin{equation}
(w^m)^T\Phi(x_i)+b^m\geq1-\xi_i, \text{if } y_i=m
\end{equation}
and
\begin{equation}
(w^m)^T\Phi(x_i)+b^m\leq-1+\xi_i, \text{if } y_i\neq m
\end{equation}
where the training data $x_i$ are mapped to a higher dimensional space by the function $\Phi$ and $C$ is the penalty parameter.
Minimizing $\frac{1}{2}(w^m)^Tw^m$ means that we would like to minimize the margin between two classes of data. When data are not linear separable, there is a penalty term $C\sum_{i=1}^{T}\xi^m_i$ which can reduce the number of training errors.
After solving equation (18) \& (19), there are k decision functions:
\begin{align*}
(w^1)^T\Phi(x)+b^1 \\
.......\\
.......\\
(w^k)^T\Phi(x)+b^k
\end{align*}
Depending upon the largest value of the decision function, the class of $x$ is
\begin{equation}
\text{class of x} \equiv argmax_{m=1,...,k}((w^m)^T\Phi(x)+b^m)
\end{equation}
\section{Experimental Setup}
\subsection{Database}
All India Radio (AIR) news reading of 10 Indian languages are used in this work. Original recordings are stored in .mp3 format with 31-32 kbps bitrate, the files are converted in .wav file with 128 kbps bitrate. Each speaker of a particular language has multiple sessions, where each session has 3-6 minutes of data. Complete database has around 40 hrs of data, with 12-16 speakers per language. The languages are: Assamese, Bengali, Hindi, Kannada, Kashmiri, Konkani, Malayalam, Odia , Tamil \& Telugu.
Database is divided into training set and test set. Separate set of speakers are taken for training set and testing set for a each language.\par
\subsection{Preprocessing}
\begin{figure}[!h]
    \centering
    \includegraphics[width=1\linewidth]{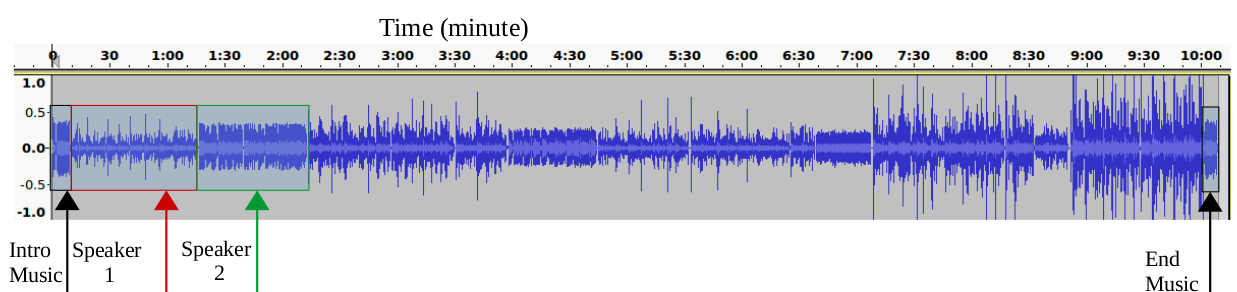}
    \caption{Example of a news reading session (language: Tamil) \label{fig:intialmp3}}
\end{figure}

The database is preprocessed to remove music and silence periods. Also some of the news recordings have multiple speakers, so it has been cleaned to make homogeneous speaker recording. In the Fig. \ref{fig:intialmp3}, we can see an example sample downloaded from the ``newsonair.nic.in'' website. It has multiple speaker recording. \par
\begin{figure}[!h]
    \centering
    \includegraphics[width=1\linewidth]{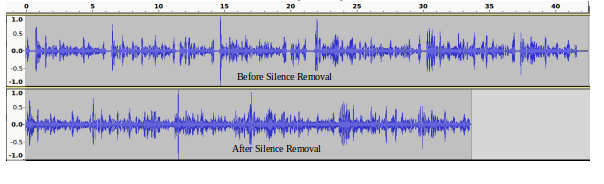}
    \caption{Before and after silence removal (language: Tamil) \label{fig:silremove}}
\end{figure}

To remove the silence period, we have used a voice activity detector system based on the energy threshold. In the Fig. \ref{fig:silremove}, we can see the speech signal before and after silence removal. Approximately 20\% silence region is there in this reading speech. 
\subsection{Training Phase}
In this work we have taken 10 Indian languages. For training the UBM we have taken total 96 minutes of data from each language. We have taken 32 audio clips of 3 minutes duration from each language. Total UBM data is of 960 minutes. We have modelled the GMM with a mixtures of 64, 128 and 256 respectively. For MAP adaptation, we have taken the same 32 audio clips of 3 minutes and split them 96 clips of 1 minute duration. After the MAP adaptation we got language, speaker and session dependent GM models.
\subsubsection{Scheme 1}
We decode the symbol sequence from this 1 minutes training files and got a sequence length of  5998$\times$1, i.e. 1 symbol per frame. The frame size is 25 ms with 10 ms of overlap. This symbol sequences have been modelled to get the n-gram models. For a mixture model of 64, we will get a bigram matrix of dimension 64$\times$64. We have transform this matrix to a vector of length 4096. Total number of training utterances are 960. So, we have got an utterance n-gram matrix of 960$\times$4096.
\subsubsection{Scheme 2}
Utterance specific mean supervector are of length 39$\times$64=2496$\times$1 for Mixture number 64. The dimension of the UBM mean supervector is also same =2496$\times$1. Total number of training utterances are 960. Hence, after taking the difference between adapted mean supervector and UBM mean, we get the difference matrix of dimension =960$\times$2496.
\subsubsection{Feature Embedding \& SVM}
To reduce the dimensionality of the utterance n-gram matrix and difference matrix, SVD is performed. After the decomposition, we got a singular value matrix of dimension 960$\times$960. We have chosen the optimal singular value energy to 55-65$\%$. After the SVD, we used linear SVM classifier to build the training models.
\subsection{Testing Phase}
\input{test_dia}
Test procedure is shown in Fig. \ref{fig:test_dia}. In the testing phase, we have used separate set of speakers. The total test data is given in table \ref{Table.one}
\begin{table}[!h]
\centering
\resizebox{0.45\textwidth}{!}{%
\begin{tabular}{|l|l|l|l|l|}
\hline
\begin{tabular}[c]{@{}l@{}}Sl.\\ No\end{tabular} & Languages & \begin{tabular}[c]{@{}l@{}}Speaker\\ No\end{tabular} & \begin{tabular}[c]{@{}l@{}}Total \\ Sessions\end{tabular} & \begin{tabular}[c]{@{}l@{}}Total \\ Duration (m)\end{tabular} \\ \hline
1.                                               & Assamese  & 6                                                    & 11                                                        & 53                                                            \\ \hline
2.                                               & Bengali   & 8                                                    & 11                                                        & 56                                                            \\ \hline
3.                                               & Hindi     & 7                                                    & 11                                                        & 54                                                            \\ \hline
4.                                               & Kannada   & 5                                                    & 12                                                        & 68                                                            \\ \hline
5.                                               & Kashmiri  & 6                                                    & 12                                                        & 29                                                            \\ \hline
6.                                               & Konkani   & 4                                                    & 9                                                         & 31                                                            \\ \hline
7.                                               & Malayalam & 8                                                    & 14                                                        & 74                                                            \\ \hline
8.                                               & Odia      & 8                                                    & 14                                                        & 68                                                            \\ \hline
9.                                               & Tamil     & 8                                                    & 12                                                        & 45                                                            \\ \hline
10.                                              & Telugu    & 6                                                    & 12                                                        & 62                                                            \\ \hline
\end{tabular}}
\caption{Test data distribution among the languages}\label{Table.one}
\end{table}
\subsubsection{Scheme 1}
 first features are extracted for a given test utterance $Y=[y_1,..y_l]$ and then these features are adapted from the UBM model, After that symbol sequence of the test utterance is decoded and test utterance n-gram vector $bv_{test}$ is estimated as follows:
\begin{equation}
SS_{t}=\operatorname*{arg\,max}_{m\in1:M} p(y_t|\lambda_{test})
\end{equation}
\begin{equation}
b_{test}(i,j)=\frac{Count(SS_{t-K}=i,SS_{t}=j)}{Count(SS_{t-K}=i)}
\end{equation}
\begin{equation}
bv_{test}=b_{test}(:)
\end{equation}
$bv_{test}$ is projected to the lower dimensional space using the same $\mathcal{L}_1$ singular values estimated during training.
\begin{equation}
bv_{test}^{\mathcal{L}_{1}}=bv_{test}{V_1}_{\mathcal{L}_{1}}{S_1}_{\mathcal{L}_{1}}^{-1}
\end{equation}
\subsubsection{Scheme 2}
Test utterance features are extracted and then UBM model is used to adapt these features. The adapted means are concatenated to get the corresponding supervector $y_{i}$. Difference vector is estimated by subtracting $y_{i}$ UBM mean supervector $U_{m}$,
\begin{equation}
\Delta y_{i}=y_{i}-U_{m}
\end{equation}
$\Delta y_{i}$ is projected to the lower dimensional space using the same $\mathcal{L}_{2}$ singular values estimated during training
\begin{equation}
\Delta y_{i}^{\mathcal{L}_{2}}=\Delta y_{i}{V_2}_{\mathcal{L}_{2}}{S_2}_{\mathcal{L}_{2}}^{-1}
\end{equation}
Finally the language of the lower dimensional test vector $\Delta y_{i}^{\mathcal{L}_{2}}$ is identified by classifying it with the trained SVMs.
\section{Results \& Discussions}
\subsection{Determining optimal energy range of singular values:}
\begin{figure}[h]
	\centering
	\begin{minipage}[b]{0.48\linewidth}
		\includegraphics[width=1\linewidth]{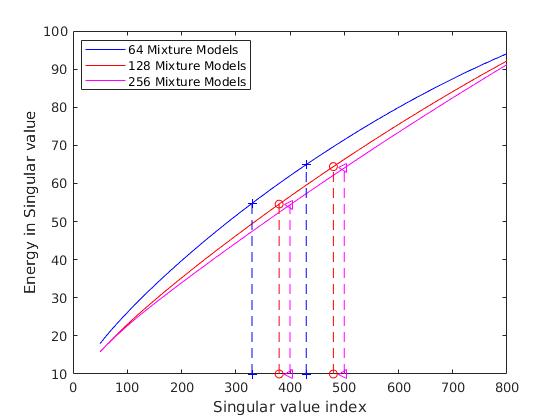}
		\centerline{(a)}
	\end{minipage}
	\begin{minipage}[b]{0.48\linewidth}
		\includegraphics[width=1\linewidth]{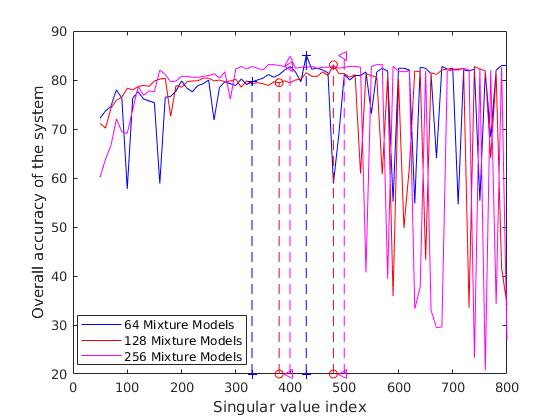}
		\centerline{(b)}		
	\end{minipage}
	\begin{minipage}[b]{0.48\linewidth}
		\includegraphics[width=1\linewidth]{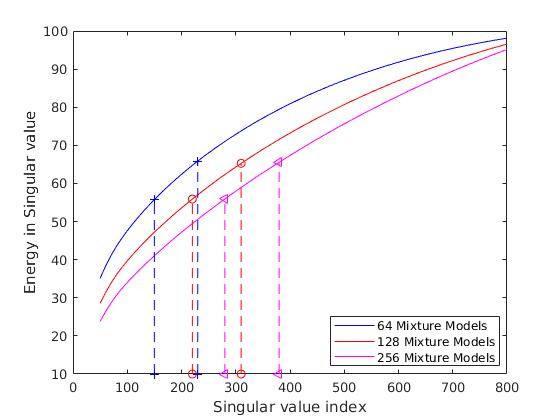}
		\centerline{(c)}
	\end{minipage}
	\begin{minipage}[b]{0.48\linewidth}
		\includegraphics[width=1\linewidth]{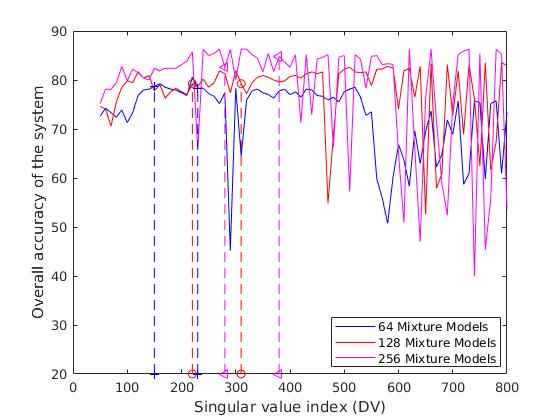}
		\centerline{(d)}		
	\end{minipage}
\caption{(a) Singular value ($S_1$) enegry distribution with respect to singular value index (b) Overall accuracy of the LID system with respect to singular value index (c) Singular value ($S_2$) enegry distribution with respect to singular value index (d) Overall accuracy of the system with respect to singular value index}\label{fig:dim_sel}
\end{figure}
We can observe from the Fig. \ref{fig:dim_sel} that for low and higher singular value dimensions the accuracy of the LID system fluctuates. We have chosen a stable accuracy area with respect to singular value energy i.e 55-65\%.
\subsection{Comparison between different n-gram modelling with overall identification of the system:}
After choosing the singular value dimension, we compared different skipgram models in terms of overall accuracy.
\begin{table*}[!h]
\centering
\resizebox{0.9\textwidth}{!}{%
\begin{tabular}{|l|l|l|l|l|l|l|l|l|l|l|l|l|l|}
\hline
                                                  &                                                            &                                                                      & \multicolumn{11}{c|}{Language Identification Accuracy (\%)}  
\\ \hline
\begin{tabular}[c]{@{}l@{}}Sl.\\ No.\end{tabular} & Modelling                                                  & \begin{tabular}[c]{@{}l@{}}Singualar\\ Value\\ Dimesion\end{tabular} & \multicolumn{1}{c|}{Assa} & \multicolumn{1}{c|}{Ben} & \multicolumn{1}{c|}{Hin} & \multicolumn{1}{c|}{Kan} & \multicolumn{1}{c|}{Kash} & \multicolumn{1}{c|}{Kon} & \multicolumn{1}{c|}{Mal} & \multicolumn{1}{c|}{Odi} & \multicolumn{1}{c|}{Tam} & \multicolumn{1}{c|}{Tel} & \multicolumn{1}{c|}{Ovl} \\ \hline
1.                                                & Bigram                                                     & 266                                                                  & 83.01                     & \textbf{98.21}                    & \textbf{100}                      & 25                       & 0                         & 96.77                    & 44.59                    & 61.76                    & 77.77                    & 50                       & 63.71                    \\ \hline
2.                                                & \begin{tabular}[c]{@{}l@{}}Skipgram\\ Shift=2\end{tabular} & 328                                                                  & \textbf{100 }                      & 89.28                    & 72.22                    & 47.05                    & 20.68                     & \textbf{100}                      & 47.29                    & 92.64                    & \textbf{95.55}                    & 69.35                    & 73.40                    \\ \hline
3.                                                & \begin{tabular}[c]{@{}l@{}}Skipgram\\ Shift=3\end{tabular} & 363                                                                  & \textbf{100 }                      & 91.07                    & 72.22                    & 45.58                    & 37.93                     & \textbf{100 }                     & 50                       & 95.58                    & \textbf{95.55}                    & 66.12                    & 75.40                    \\ \hline
4.                                                & \begin{tabular}[c]{@{}l@{}}Skipgram\\ Shift=4\end{tabular} & 384                                                                  & 98.11                     & 96.42                    & 59.25                    & 45.58                    & 55.17                     & \textbf{100}                      & 62.16                    & \textbf{98.52}                    & \textbf{95.55 }                   & 64.51                    & 77.52                    \\ \hline
5.                                                & \begin{tabular}[c]{@{}l@{}}Skipgram\\ Shift=5\end{tabular} & 400                                                                  & 94.33                     & 87.5                     & 74.07                    & 50                       & 72.41                     & \textbf{100  }                    & 64.86                    & \textbf{98.52}                    & 93.33                    & 74.19                    & 80.92                    \\ \hline
6.                                                & \begin{tabular}[c]{@{}l@{}}Skipgram\\ Shift=6\end{tabular} & 408                                                                  & 96.22                     & 87.5                     & 74.07                    & 41.17                    & \textbf{79.31}                     & \textbf{100}                      & \textbf{94.59 }                   & 94.11                    & 86.66                    & 56.45                    & 81.01                    \\ \hline
7.                                                & \begin{tabular}[c]{@{}l@{}}Skipgram\\ Shift=7\end{tabular} & 425                                                                  & 96.22                     & 91.07                    & 83.33                    & \textbf{63.29}                    & \textbf{79.31}                     & \textbf{100                     } & 75.67                    & 89.70                    & 93.33                    & \textbf{79.03}                    & \textbf{85.09}                    \\ \hline
\end{tabular}}
\caption{Comparison between different sparse skipgram modelling, Mix=64, Duration=30 sec, Singular value energy 60-65\%}\label{Table.2}
\end{table*}
In Table \ref{Table.2}, we can see the overall accuracy of different n-gram models. We can see that the overall accuracy has increased with increasing shift parameter. We can observe that for bigram, performance of some languages are showing very good identification accuracy (above 95\%), such as, Hindi, Bengali, Konkani but languages like Kashmiri, Kannada, Malayalam have shown very poor accuracy. As we gradually increased the shift, the accuracy of Kashmiri, Kannada, Malayalam and Telugu has improved. Observations are listed below:
\begin{enumerate}
\item Assamese language has good identification accuracy irrespective of the shift. Although, at shift=2 and 3, it is showing 100\% result.

\item Bengali language is showing similar characteristic like Assamese but the best accuracy is with bigram model
.
\item Hindi also shows best accuracy with bigram model. Accuracy is lowest at shift=4, where Hindi language has misclassified mostly as Bengali and Kannada. The accuracy has increased after that and reached above 90\%.

\item Kannada Language has poor accuracy with bigram and most skipgram models. It achieved a moderate identification accuracy at shift=7. Kannada language is mostly misclassified as Tamil and Hindi.

\item Kashmiri language has no identification at bigram and its identification rate has increased with increament of the shift parameter. Kashmiri language has mostly misclassified as Odia.

\item Konkani language has overall good accuracy irrespective of the shift parameter.

\item Malayalam language has shown best accuracy at shift =6. Accuracy is poor at low shift.

\item Odia language has shown good accuracy except in bigram modelling. In bigram modelling, Odia language has  misclassified as Hindi and Bengali.

\item Tamil has overall good accuracy except at bigram modelling.

\item Telugu has shown gradual increment in overall identification accuracy with increase in shift parameter.        
 
\end{enumerate}
\subsection{Overall accuracy with different mixture components:}
\subsubsection{Scheme 1}
From Table \ref{Table.3}, we can see that at test duration of 30 second, UBM with 64 mixtures gives better overall LID accuracy in less projected dimension. Although, the identification accuracy of Hindi, Tamil and Telugu language are better with 256 mixtures but the identification efficiencies of Kannada and Malayalam are better with 64 mixtures. We can also see that the accuracies of Bengali, Hindi, Kashmiri and Telugu languages are better with higher mixture components. Kannada has lowest identification rate for all the mixture components and it seems that the skipgram model is failed to capture language specific information for Kannada language.      
\begin{table*}[!h]
\centering
\resizebox{1\textwidth}{!}{%
\begin{tabular}{|l|l|l|l|l|l|l|l|l|l|l|l|l|l|l|l|}
\hline
\multirow{2}{*}{\begin{tabular}[c]{@{}l@{}}Sl.\\ No.\end{tabular}} & \multirow{2}{*}{\begin{tabular}[c]{@{}l@{}}No. of \\ Mixtures\end{tabular}} & \multicolumn{1}{c|}{\multirow{2}{*}{Model}}                & \multirow{2}{*}{\begin{tabular}[c]{@{}l@{}}Singular\\ Value Dim\end{tabular}} & \multicolumn{1}{c|}{\multirow{2}{*}{\begin{tabular}[c]{@{}c@{}}Test\\ Duration (sec)\end{tabular}}} & \multicolumn{11}{c|}{Overall accuracy (\%)}                                             \\ \cline{6-16} 
                                                                   &                                                                             & \multicolumn{1}{c|}{}                                      &                                                                                       & \multicolumn{1}{c|}{}                                                                                 & Assa  & Ben   & Hin   & Kan   & Kash  & Kon & Mal   & Odi   & Tam   & Tel   & Ovl   \\ \hline
1.                                                                 & 64                                                                          & \begin{tabular}[c]{@{}l@{}}Skipgram\\ Shift=7\end{tabular} & 425                                                                                   & 30                                                                                                    & 96.22 & 91.07 & 83.33 & 63.29 & 79.31 & 100 & 75.67 & 89.70 & 93.33 & 79.03 & 85.09 \\ \hline
2.                                                                 & 128                                                                         & \begin{tabular}[c]{@{}l@{}}Skipgram\\ Shift=7\end{tabular} & 480                                                                                   & 30                                                                                                    & 96.22 & 94.64 & 70.37 & 61.76 & 79.31 & 100 & 66.21 & 91.17 & 95.55 & 75.80 & 83.10 \\ \hline
3.                                                                 & 256                                                                         & \begin{tabular}[c]{@{}l@{}}Skipgram\\ Shift=7\end{tabular} & 500                                                                                   & 30                                                                                                    & 96.22 & 92.85 & 88.88 & 51.47 & 79.31 & 100 & 58.10 & 91.17 & 97.77 & 93.54 & 84.9  \\ \hline
\end{tabular}}
\caption{Overall LID accuracy with different mixture components (Scheme:1)}\label{Table.3}
\end{table*}
\subsubsection{Scheme 2} 
\begin{table*}[!h]
\centering
\resizebox{1\textwidth}{!}{%
\begin{tabular}{|l|l|l|l|l|l|l|l|l|l|l|l|l|l|l|}
\hline
\multirow{2}{*}{\begin{tabular}[c]{@{}l@{}}Sl.\\ No\end{tabular}} & \multirow{2}{*}{\begin{tabular}[c]{@{}l@{}}Mixture \\ No\end{tabular}} & \multirow{2}{*}{\begin{tabular}[c]{@{}l@{}}Singular\\ Value Dim\end{tabular}} & \multirow{2}{*}{\begin{tabular}[c]{@{}l@{}}Test\\ Duration (Sec)\end{tabular}} & \multicolumn{11}{c|}{Overall accuracy (\%)}                                             \\ \cline{5-15} 
                                                                  &                                                                        &                                                                               &                                                                                & Assa  & Ben   & Hin   & Kan   & Kash  & Kon & Mal   & Odi   & Tam   & Tel   & Ovl   \\ \hline
1.                                                                & 64                                                                     & 220                                                                           & 30                                                                             & 88.67 & 85.71 & 85.18 & 61.76 & 58.62 & 100 & 63.51 & 82.35 & 97.77 & 82.25 & 80.58 \\ \hline
2.                                                                & 128                                                                    & 300                                                                           & 30                                                                             & 88.67 & 83.92 & 87.03 & 72.05 & 72.41 & 100 & 58.10 & 94.11 & 88.88 & 74.19 & 81.9  \\ \hline
3.                                                                & 256                                                                    & 360                                                                           & 30                                                                             & 96.22 & 80.35 & 87.03 & 79.41 & 75.86 & 100 & 62.16 & 85.29 & 97.77 & 90.32 & 85.4  \\ \hline
\end{tabular}}
\caption{Overall LID performance with different mixture components(Scheme:2)}\label{Table.7}
\end{table*}
In Table. \ref{Table.7}, we can see that at a fixed text duration, the overall performance has significantly increased with the increase in mixture components. We can see that the performance of Assamese, Kannada and Telugu has increased at 256 mixture components. We can infer that the higher mixture components captures more language specific information. 
\subsection{Overall accuracy at different test durations:}
\subsubsection{Scheme 1}
\begin{figure}[h]
	\centering
	\begin{minipage}[b]{0.48\linewidth}
		\includegraphics[width=1\linewidth]{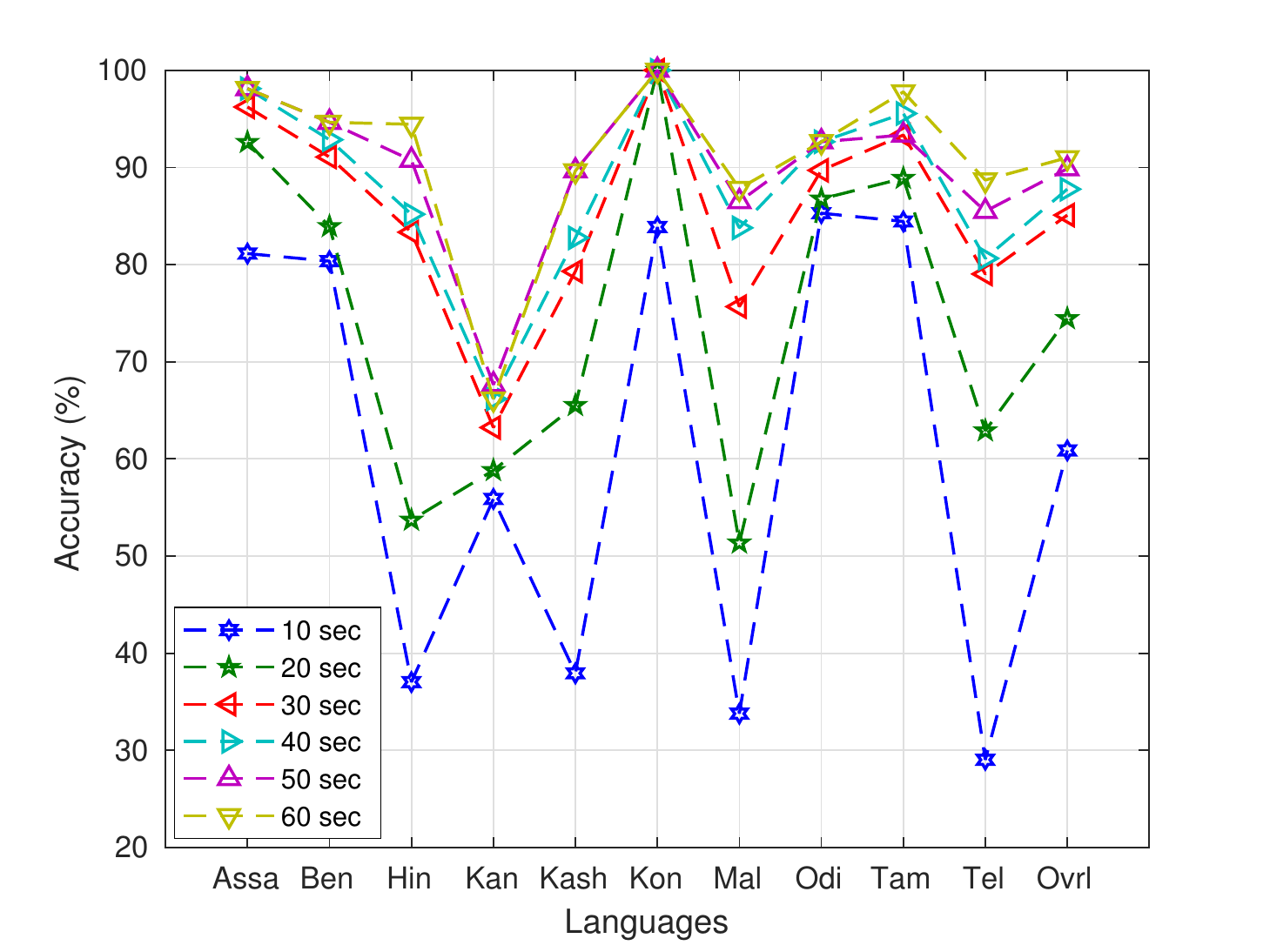}
		\centerline{(a)}
	\end{minipage}
		\begin{minipage}[b]{0.48\linewidth}
		\includegraphics[width=1\linewidth]{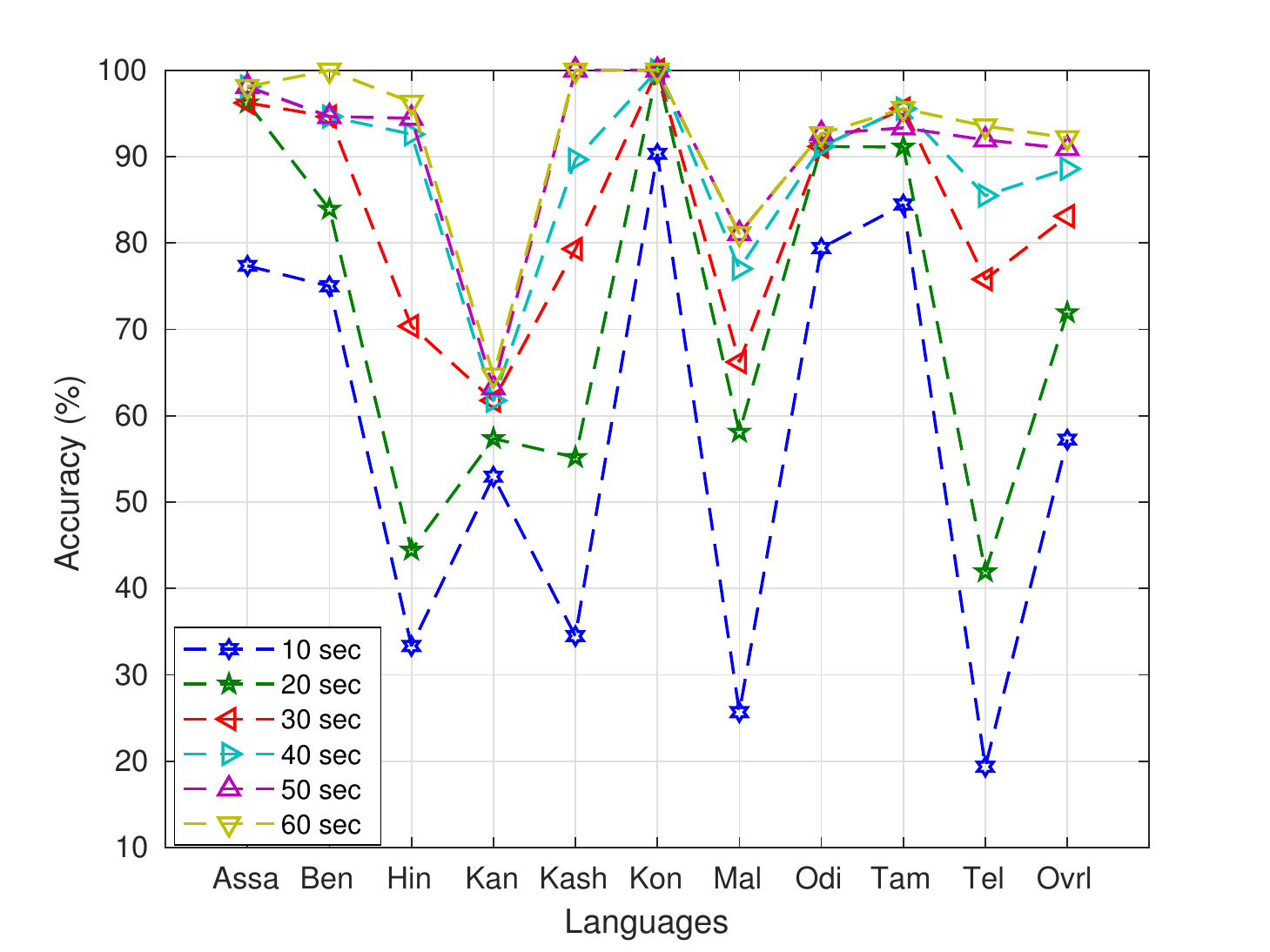}
		\centerline{(b)}		
	\end{minipage}
	\begin{minipage}[b]{0.48\linewidth}
		\includegraphics[width=1\linewidth]{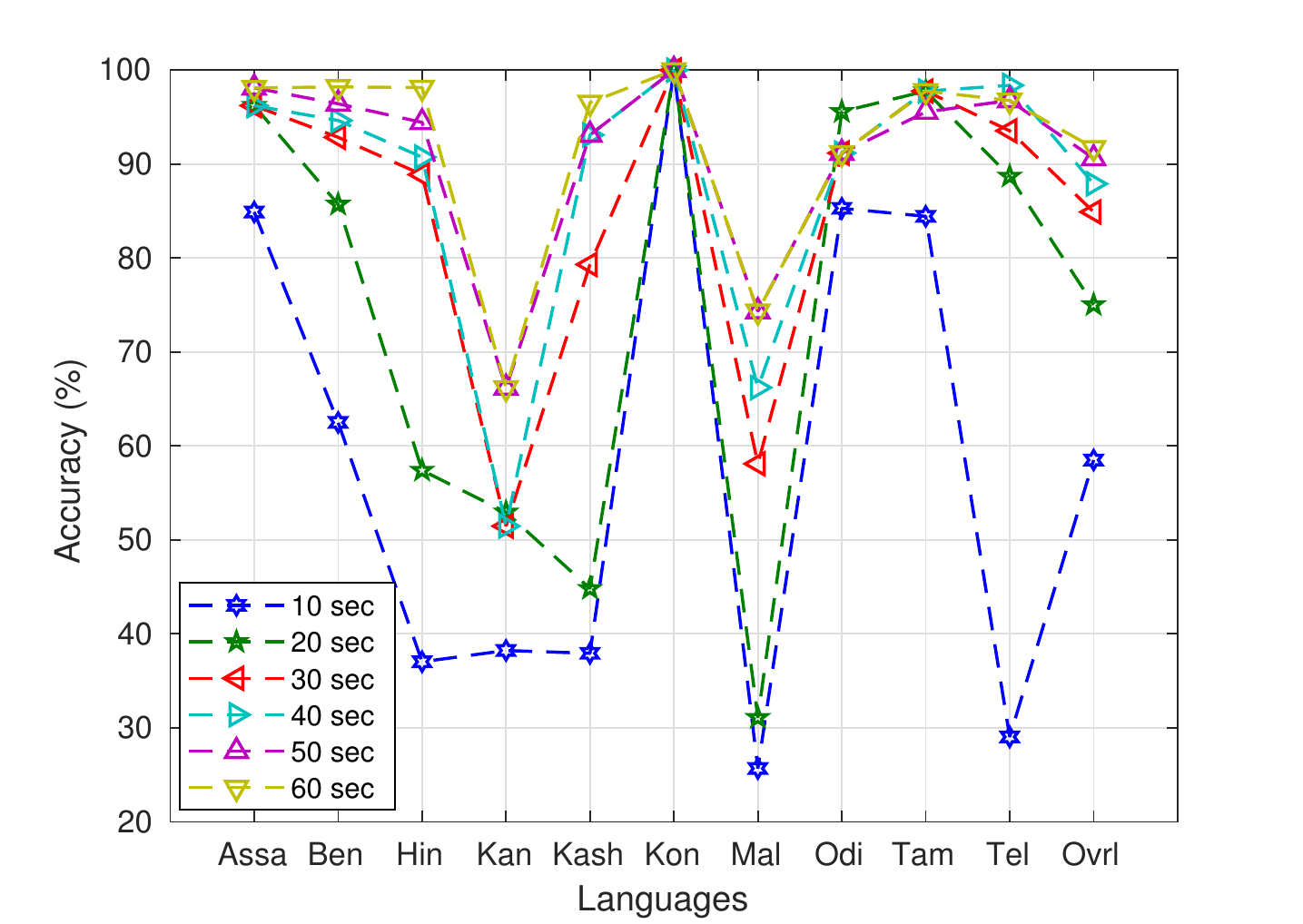}
		\centerline{(c)}		
	\end{minipage}
	\caption{Comparison of performance of scheme-1 at different test duration with (a) 64 Mixtures (b) 128 Mixtures (c) 256 Mixtures}
	\label{fig:test_dur_sc1}
\end{figure}

performance of LID system for different test duration and different mixture components are shown in Fig. \ref{fig:test_dur_sc1}. We can observe from the figure that the overall accuracy is good for test duration 10, 20 \& 30 seconds with 64 mixture components but for test duration $>$ 30 seconds, 128 \& 256 mixture components performed better.
\subsubsection{Scheme 2}
\begin{figure}[h]
	\centering
	\begin{minipage}[b]{0.48\linewidth}
		\includegraphics[width=1\linewidth]{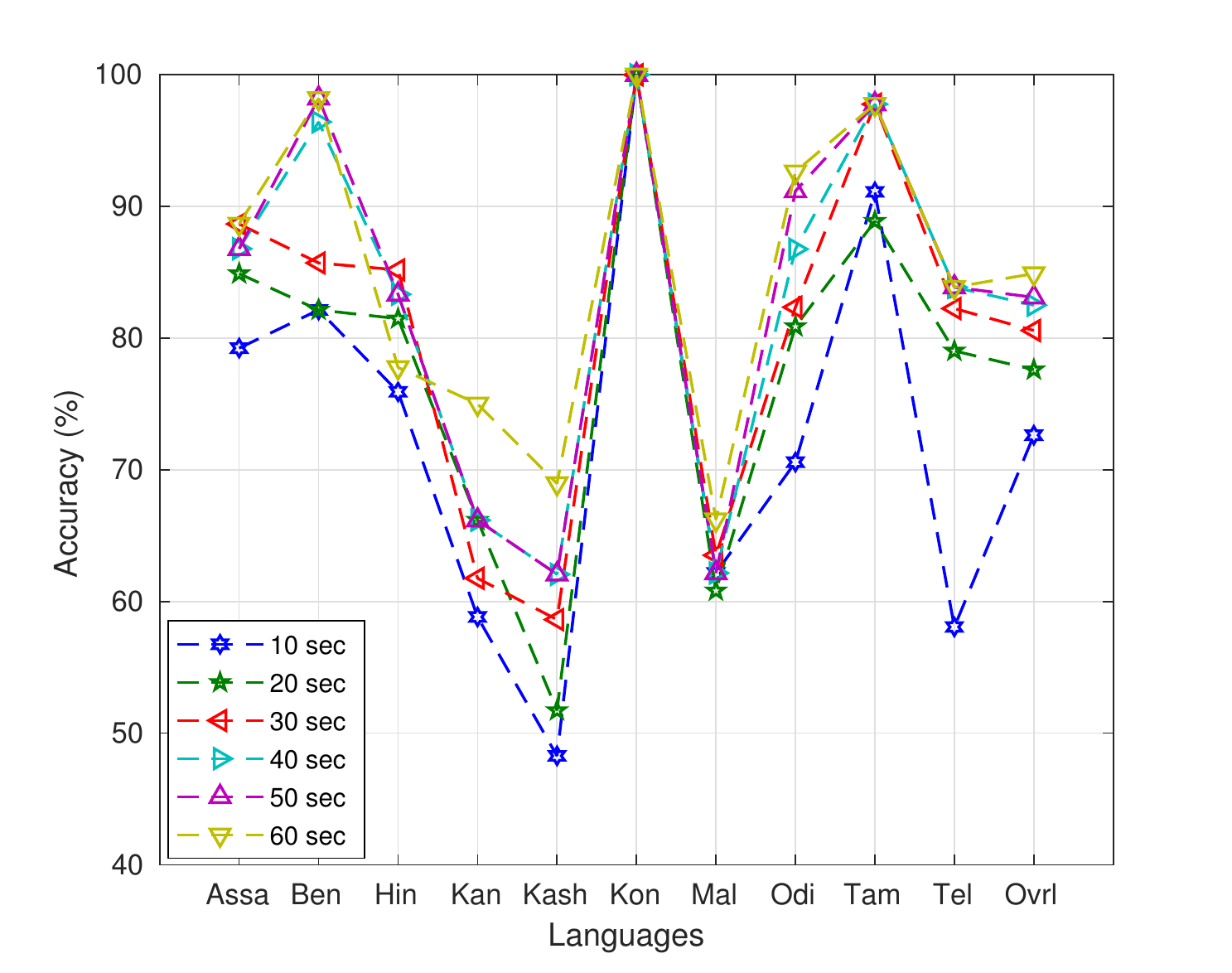}
		\centerline{(a)}
	\end{minipage}
		\begin{minipage}[b]{0.48\linewidth}
		\includegraphics[width=1.1\linewidth]{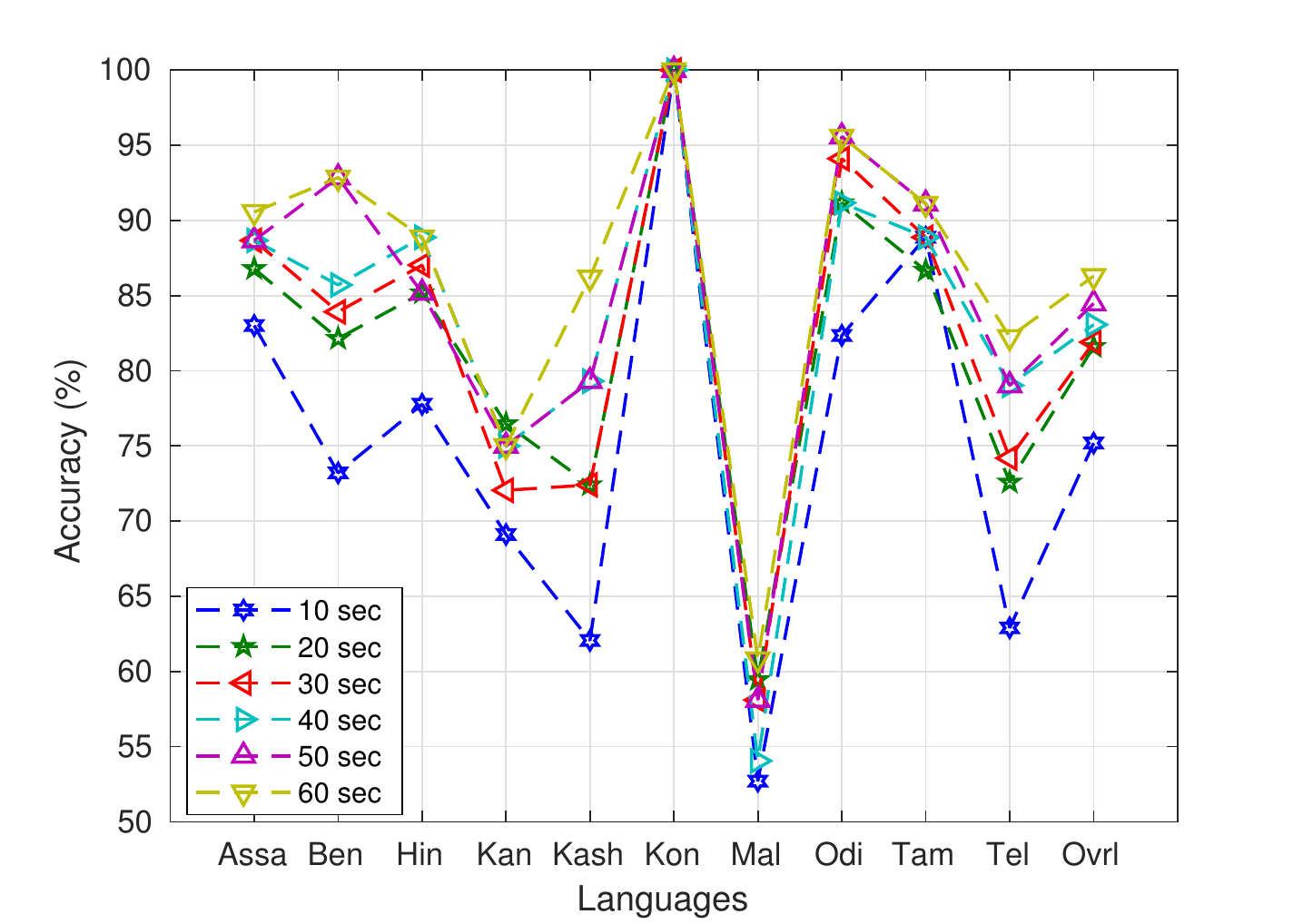}
		\centerline{(b)}		
	\end{minipage}
	\begin{minipage}[b]{0.48\linewidth}
		\includegraphics[width=1\linewidth]{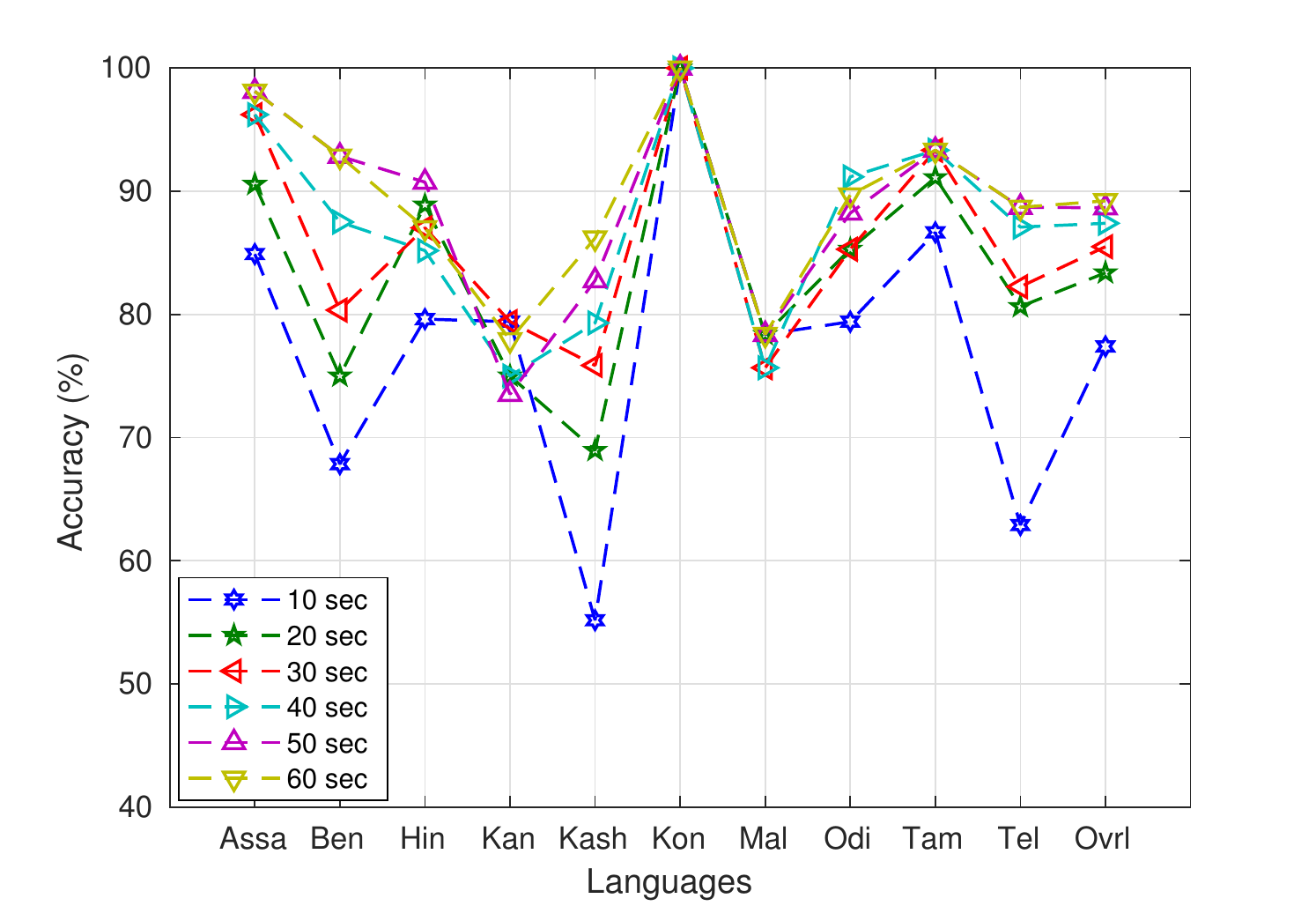}
		\centerline{(c)}		
	\end{minipage}
	\caption{Comparison of performance of scheme-2 at different test duration with (a) 64 Mixtures (b) 128 Mixtures (c) 256 Mixtures}
	\label{fig:test_dur_sc2}
\end{figure}

Performance of LID system for different test duration and different mixture components are shown in Fig. \ref{fig:test_dur_sc2}. We can observe from the figures that the overall accuracy is better at all the test duration with 256 mixture components. Also, performance of all the languages improved with 256 mixture components.

\subsection{Comparison between scheme 1 \& 2}
\begin{figure}[h]
	\centering
	\begin{minipage}[b]{0.48\linewidth}
		\includegraphics[width=1\linewidth]{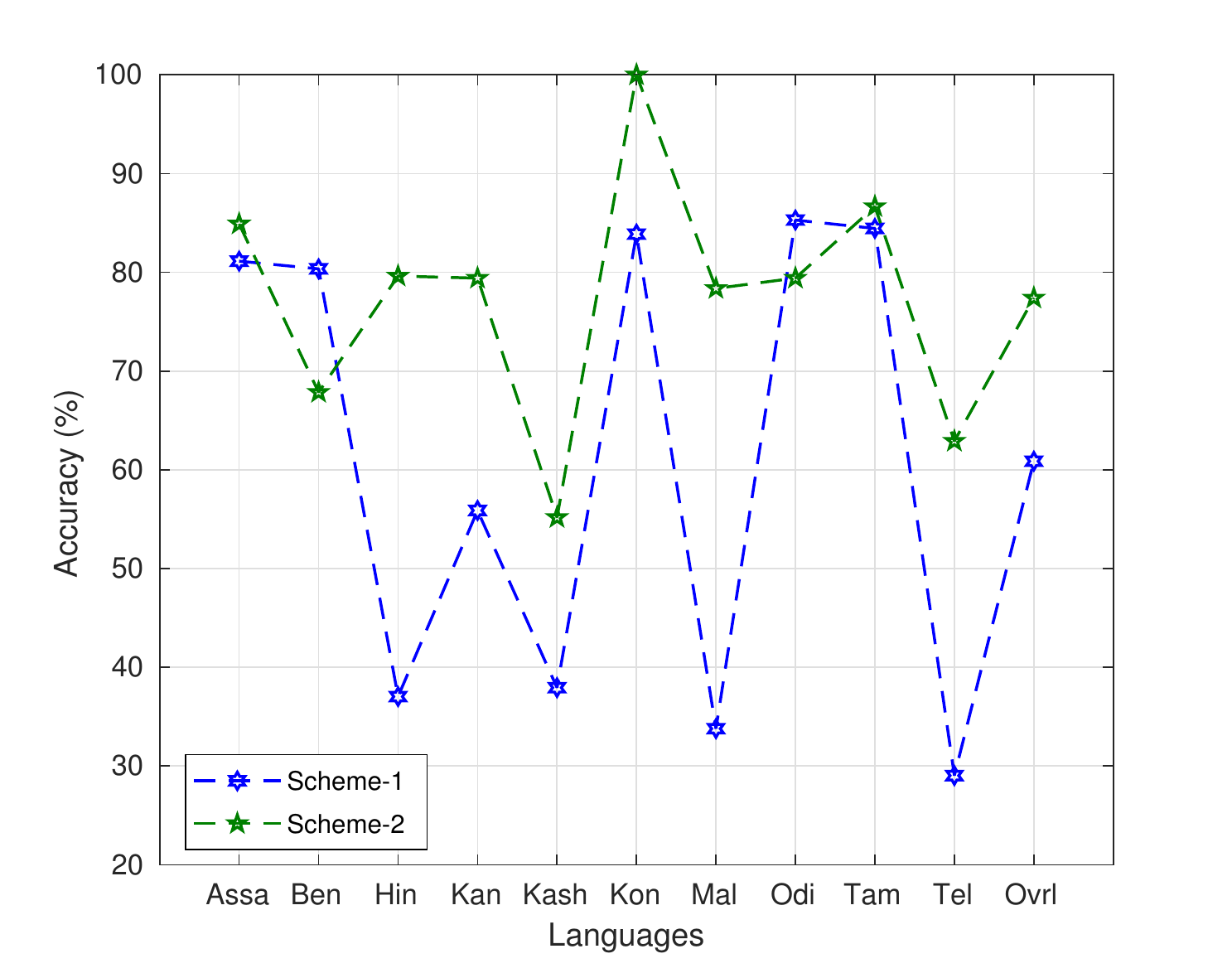}
		\centerline{(a)}
	\end{minipage}
	\begin{minipage}[b]{0.48\linewidth}
		\includegraphics[width=1\linewidth]{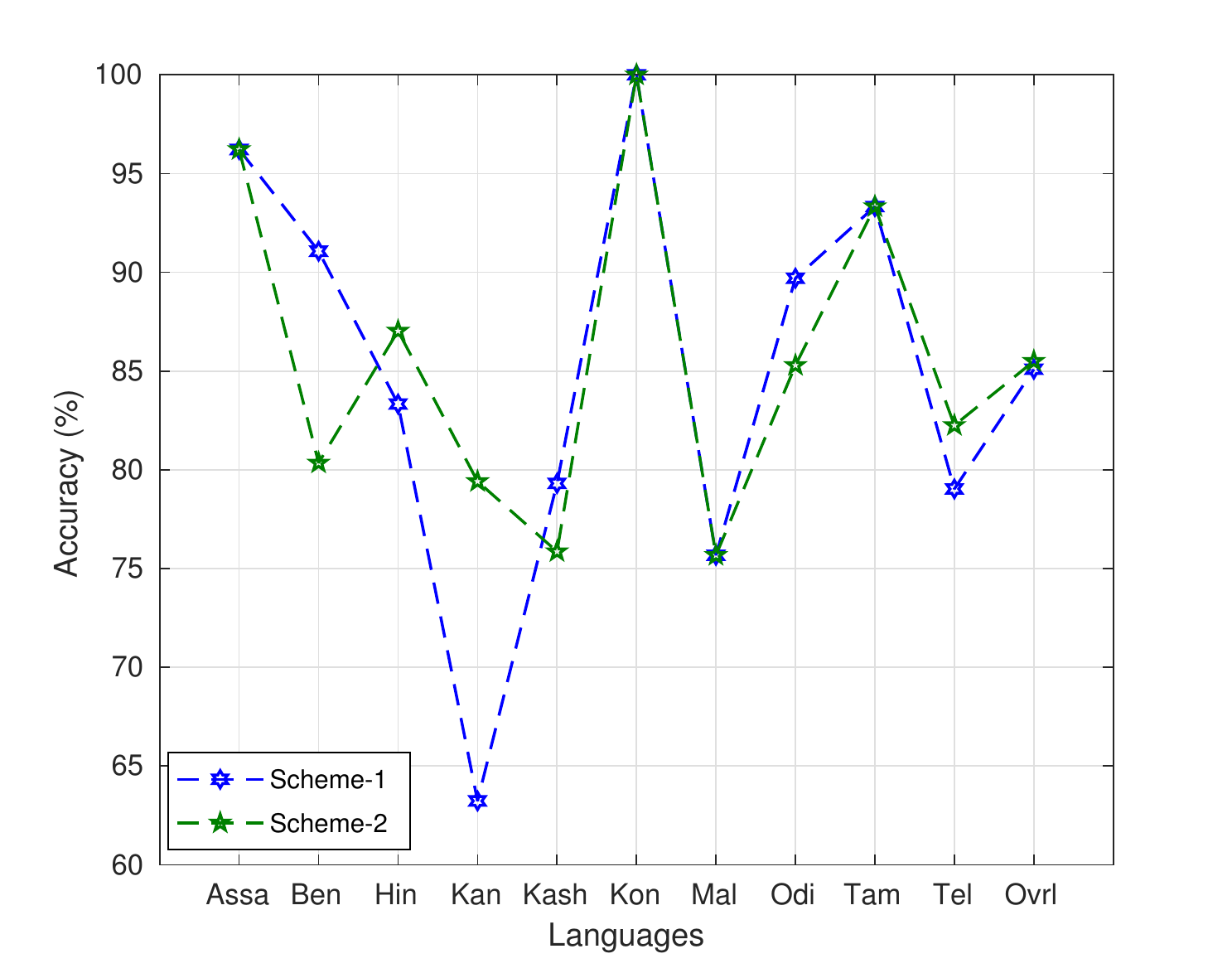}
		\centerline{(b)}		
	\end{minipage}
		\begin{minipage}[b]{0.48\linewidth}
		\includegraphics[width=1\linewidth]{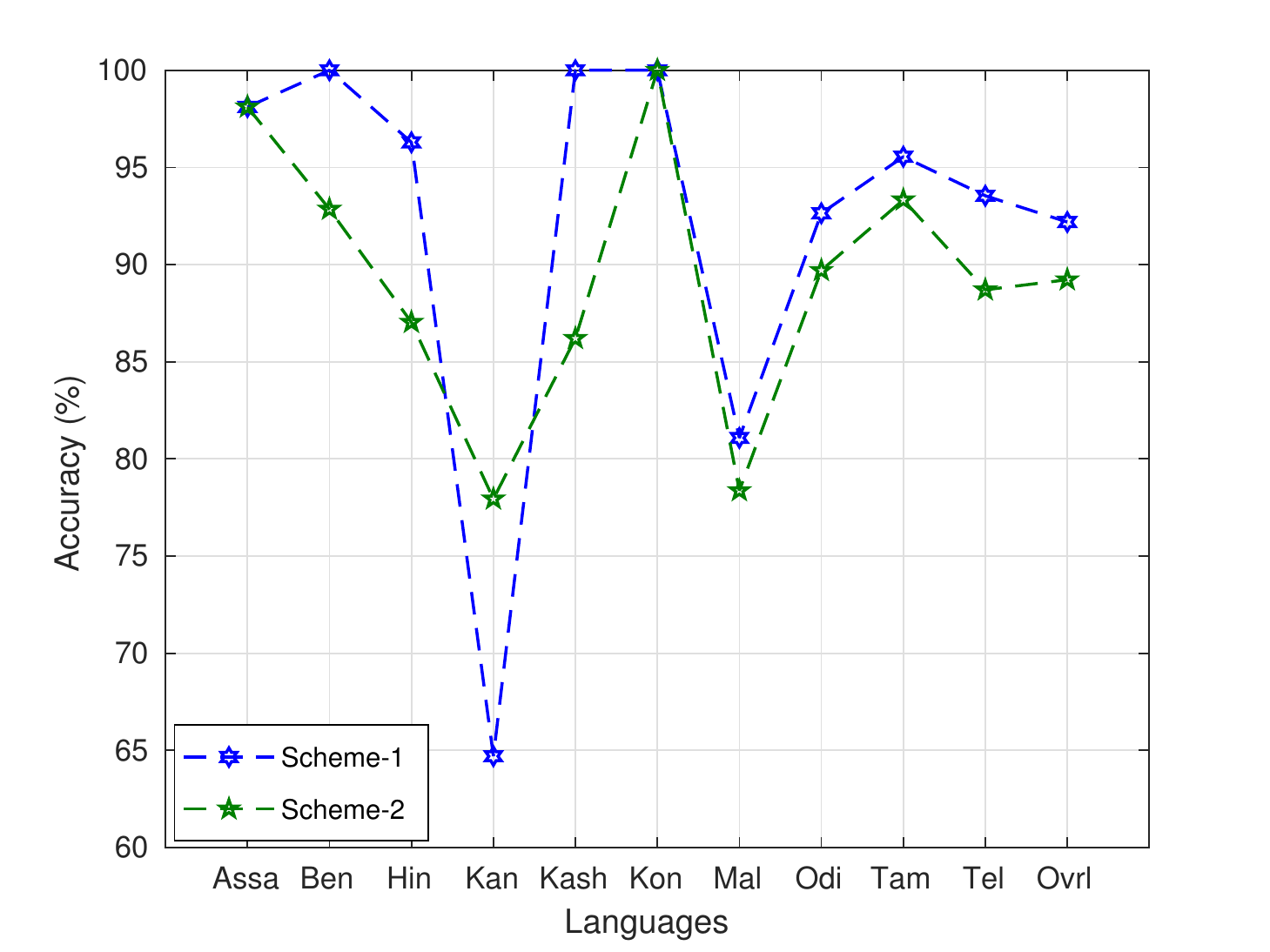}
		\centerline{(c)}		
	\end{minipage}
\caption{Comparison between scheme 1 \& 2 at tes duration (a) 10 sec (b) 30 sec (c) 60 sec}\label{fig:comp_10_20_30}
\end{figure}

In Fig. \ref{fig:comp_10_20_30}, We can see the comparison of the above mentioned two schemes at three test duration: 10, 30 \& 60 seconds. 
The overall performance of scheme-1 is poor compared to scheme-2 but the performance has improved at test duration 30 and 60 seconds. Identification accuracy of Kannada language is poor in scheme-1 for all the test duration. Whereas, Bengali, Odia, Tamil have better accuracy in scheme-1 than scheme-2 for 30 and 60 seconds test duration.

\subsection{Language based speech segmentation}
\input{lang_dia}
For language based speech segmentation, we have taken a set of four languages from the ten language group and make different sub-groups. We have taken four language subgroups by considering that at a time, four different languages can come in a conversation.
The sub-groups are listed below:
\begin{itemize}
\item Sub-group-I: Bengali, Odia, Tamil \& Malayalam.
\item Sub-group-II: Assamese, Bengali, Kannada \& Telugu.
\item Sub-group-III: Kannada, Telugu, Malayalam \& Tamil.
\item Sub-gropu-IV: Hindi, Kashmiri, Konkani \& Odia. 
\end{itemize}
In the first two subgroups (I and II), two languages are Indo-Aryan languages and other two are Dravidian languages. In the third sub-group all languages are Dravidian languages. In the last group all the languages are Indo-Aryan languages.\par
We have synthetically concatenated languages in each subgroups, Each of the languages has minimum 3 segments and the segment duration is in between 6-30 seconds, 2-5 seconds sliding window is used with 1 second shift. We have taken ten language training model to evaluate the results.
We have estimated the segmentation accuracy with both the schemes 1 \& 2. Results are listed below in tabular form.
\begin{table}[!h]
\centering
\begin{tabular}{|l|l|l|l|}
\hline
Sl.No. & \multicolumn{1}{c|}{\begin{tabular}[c]{@{}c@{}}Slide\\ Window Duration (SWD) (sec)\end{tabular}} & \multicolumn{1}{c|}{Scheme-1} & \multicolumn{1}{c|}{Scheme-2} \\ \hline
1. & 2 & 50.63 & 56.36 \\ \hline
2. & 3 & 56.05 & 61.14 \\ \hline
3. & 4 & 60.19 & 62.73 \\ \hline
4. & 5 & 64.37 & 65.56 \\ \hline
\end{tabular}
\caption{Language segmentation accuracy with sub-group-I languages}\label{subgroup-I}
\end{table} 
\begin{table}[!h]
\centering
\begin{tabular}{|l|l|l|l|}
\hline
Sl.No. & \multicolumn{1}{c|}{\begin{tabular}[c]{@{}c@{}}Slide\\ Window Duration (SWD) (sec)\end{tabular}} & \multicolumn{1}{c|}{Scheme-1} & \multicolumn{1}{c|}{Scheme-2} \\ \hline
1. & 2 & 55.83 & 63.73 \\ \hline
2. & 3 & 57.09 & 66.02 \\ \hline
3. & 4 & 58.2 & 66.07 \\ \hline
4. & 5 & 58.67 & 66.82 \\ \hline
\end{tabular}
\caption{Language segmentation accuracy with sub-group-II languages}\label{subgroup-II}
\end{table} 
\begin{table}[!h]
\centering
\begin{tabular}{|l|l|l|l|}
\hline
Sl.No. & \multicolumn{1}{c|}{\begin{tabular}[c]{@{}c@{}}Slide\\ Window Duration (SWD) (sec)\end{tabular}} & \multicolumn{1}{c|}{Scheme-1} & \multicolumn{1}{c|}{Scheme-2} \\ \hline
1. & 2 & 48.99 & 58.77 \\ \hline
2. & 3 & 50.53 & 61.13 \\ \hline
3. & 4 & 54.3 & 60.77 \\ \hline
4. & 5 & 56.77 & 59.36 \\ \hline
\end{tabular}
\caption{Language segmentation accuracy with sub-group-III languages}\label{subgroup-III}
\end{table}
\begin{table}[!h]
\centering
\begin{tabular}{|l|l|l|l|}
\hline
Sl.No. & \multicolumn{1}{c|}{\begin{tabular}[c]{@{}c@{}}Slide\\ Window Duration (SWD) (sec)\end{tabular}} & \multicolumn{1}{c|}{Scheme-1} & \multicolumn{1}{c|}{Scheme-2} \\ \hline
1. & 2 & 62.45 & 76.27 \\ \hline
2. & 3 & 64.23 & 78.13 \\ \hline
3. & 4 & 65.82 & 79.77 \\ \hline
4. & 5 & 67.27 & 78.27 \\ \hline
\end{tabular}
\caption{Language segmentation accuracy with sub-group-IV languages}\label{subgroup-IV}
\end{table}
We can see from the Table \ref{subgroup-I}, scheme-2 has outperformed scheme-1. At SWD=5 sec, the performance of the two approaches is comparable.

\begin{figure}[h]
	\centering
	\begin{minipage}[b]{0.48\linewidth}
		\includegraphics[width=1\linewidth]{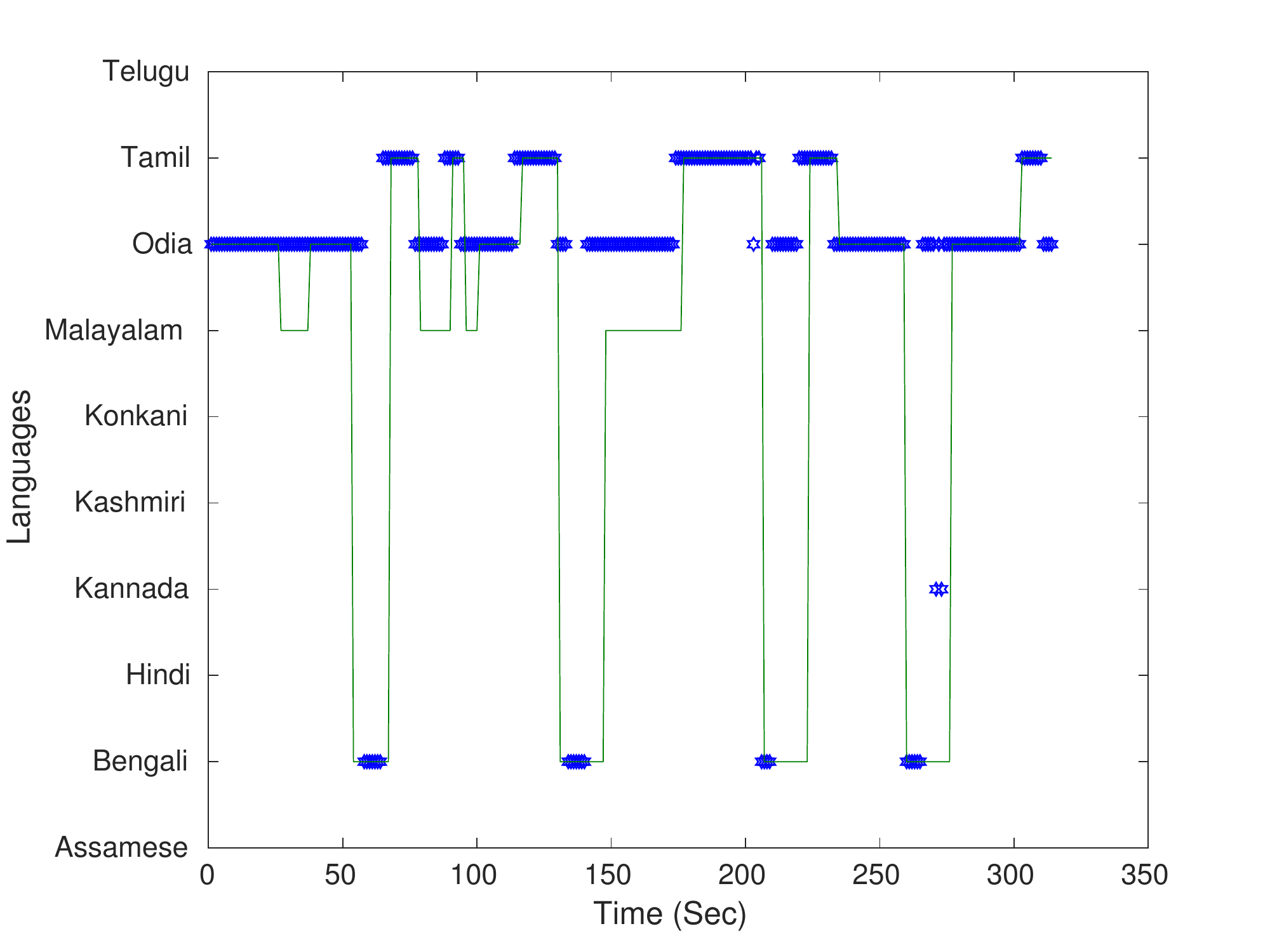}
		\centerline{(a)}
	\end{minipage}
	\begin{minipage}[b]{0.48\linewidth}
		\includegraphics[width=1\linewidth]{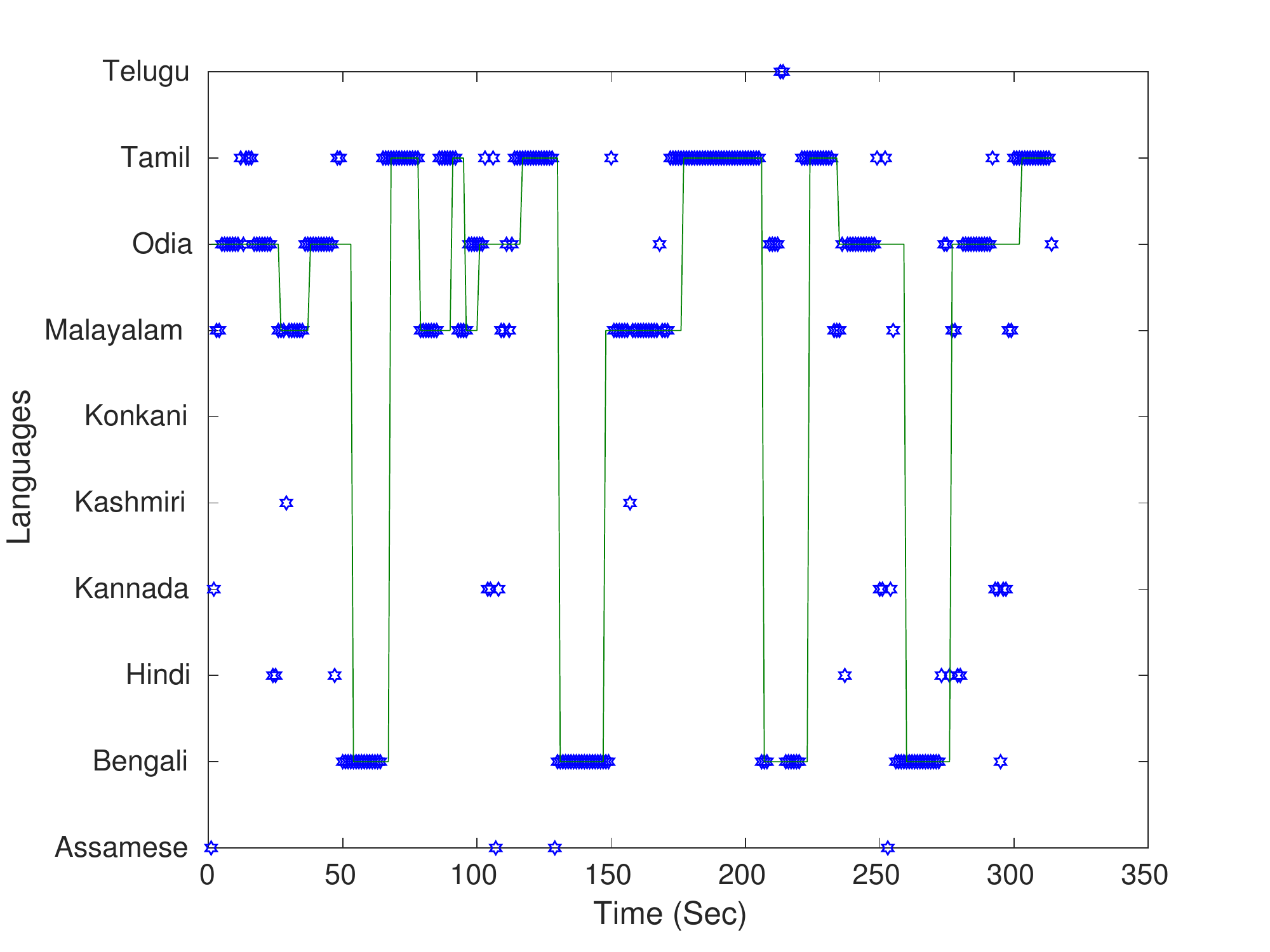}
		\centerline{(b)}		
	\end{minipage}
\caption{Language based speech segmentation with (a) Scheme-1 (b) Scheme-2 (Green solid line is showing the ground truth language and the blue marker is showing the predicted language)}\label{fig:lang_dia_Set1}
\end{figure}
In Fig. \ref{fig:lang_dia_Set1}, we can see that the scheme-1 has poor identification of Malayalam language. Most of the time the Malayalam language is identified as Odia language. But in the case of scheme-2, it has identified Malayalam language better. From Table \ref{subgroup-I}, we can see that scheme-2, has better segmentation accuracy than scheme-1 at all SWD. The identification of Odia and Tamil language is better in scheme-1 than scheme-2.\par    
\begin{figure}[h]
	\centering
	\begin{minipage}[b]{0.48\linewidth}
		\includegraphics[width=1\linewidth]{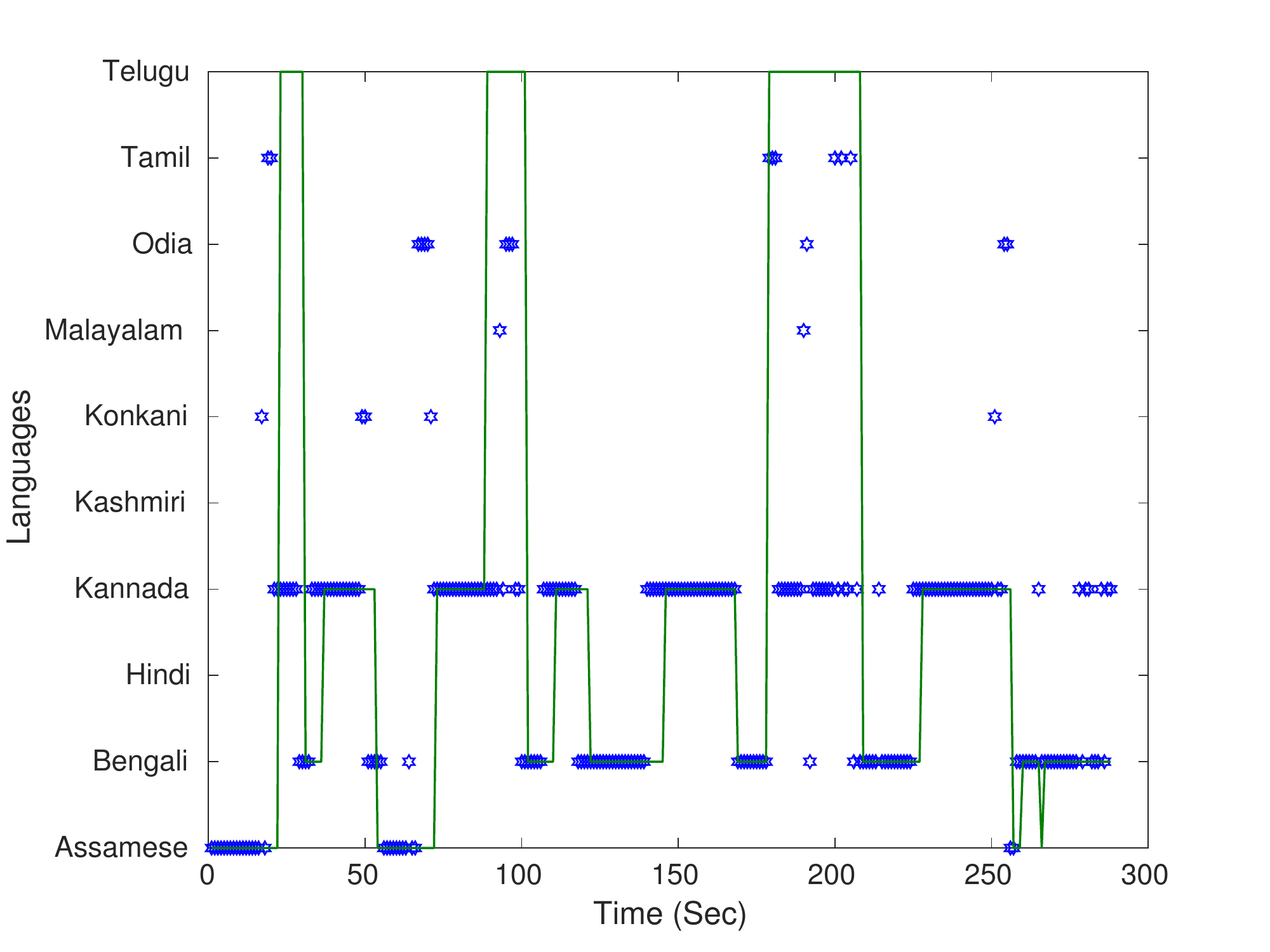}
		\centerline{(a)}
	\end{minipage}
	\begin{minipage}[b]{0.48\linewidth}
		\includegraphics[width=1\linewidth]{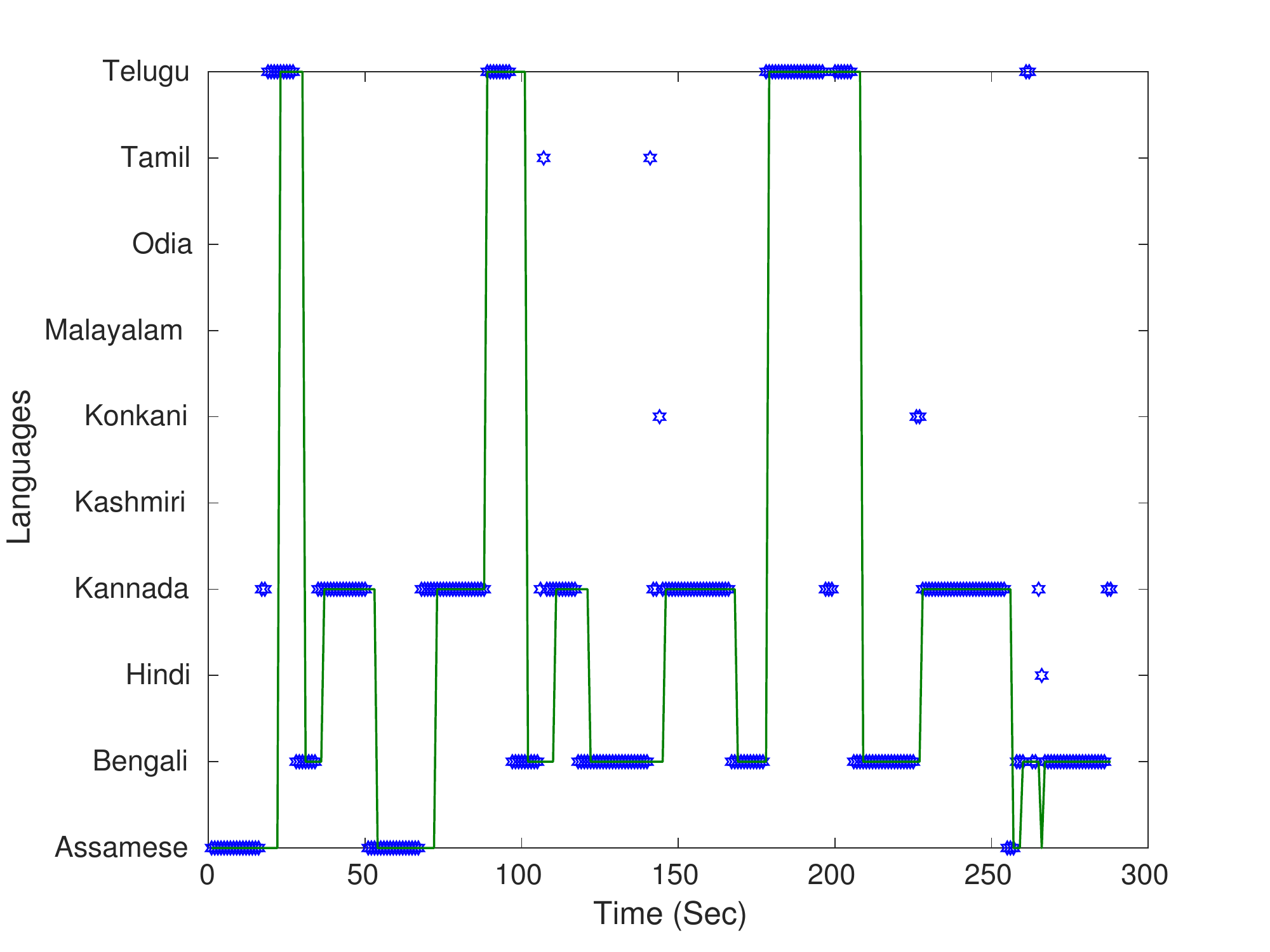}
		\centerline{(b)}		
	\end{minipage}
\caption{Language based speech segmentation with (a) Scheme-1 (b) Scheme-2 (Green solid line is showing the ground truth language and the blue marker is showing the predicted language)}\label{fig:lang_dia_Set2}
\end{figure}
In Fig. \ref{fig:lang_dia_Set2}, we can see that scheme-1 has misclassified Telugu as Kannada, as they are similar languages. Scheme-2 has distinguished Telugu and Kannada better. But, both of the schemes identified Assamese and Bengali language well. From Table \ref{subgroup-II}, we can see that, scheme-2 has outperformed scheme-1.\par
Similarly for Table \ref{subgroup-III} \& \ref{subgroup-IV}, we can see that language segmentation accuracy is better at scheme-2 than scheme-1. In Sub-group-III, the scheme-1 has poor segmentation accuracy than the scheme-2 because in scheme-1, the Telugu language is often misclassified as Kannada. In subgroup-IV, the segmentation accuracy of scheme-2 is better among the other subgroups. Here the misclassification occurs between Hindi and Odia language. Altogether we can say that as the sliding window size is short, and the performance of scheme-2 is better at short duration test speech, that is the reason of performance improvement of scheme-2.\par
\begin{figure}[h]
	\centering
	\begin{minipage}[b]{0.48\linewidth}
		\includegraphics[width=1\linewidth]{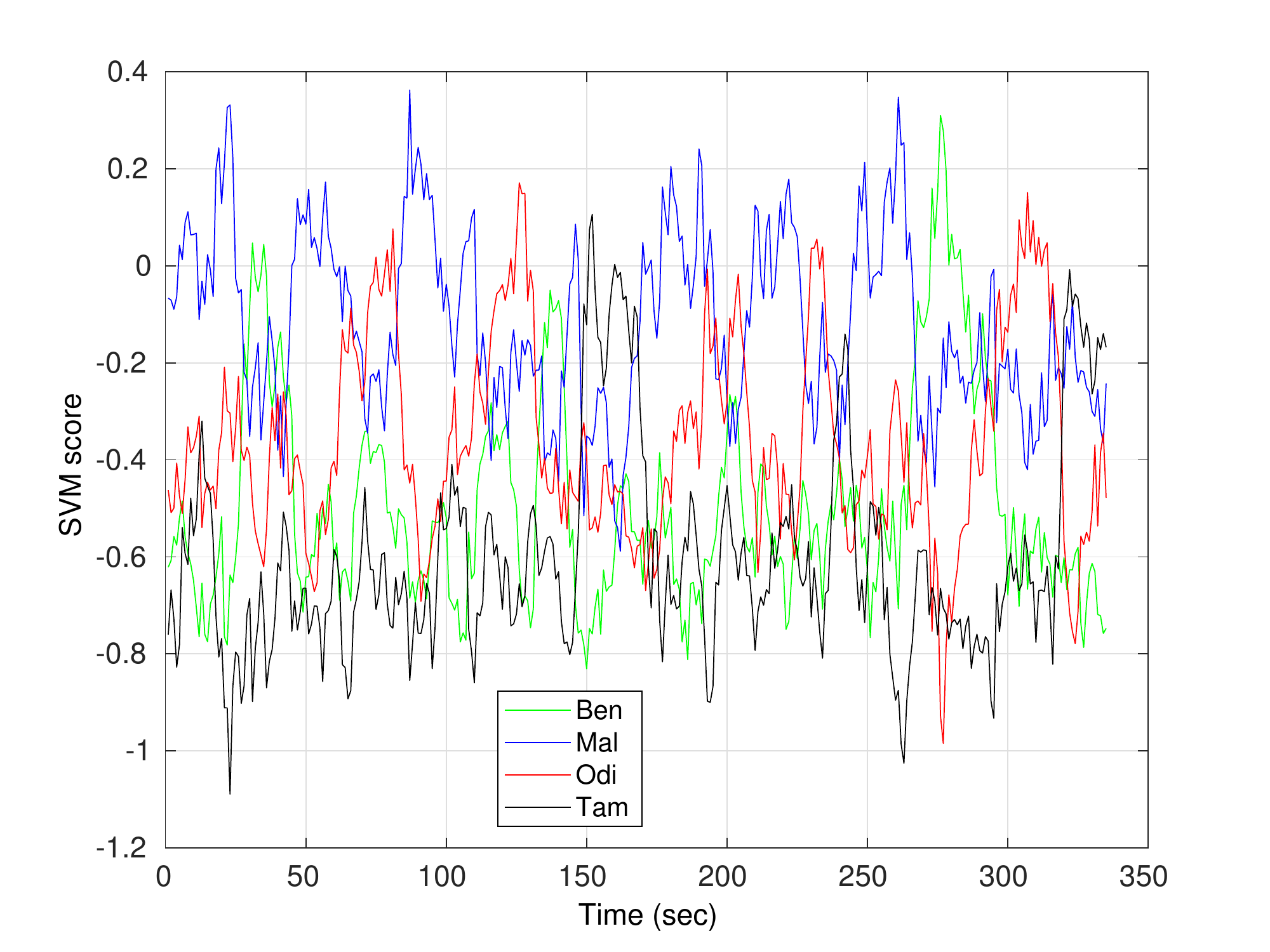}
		\centerline{(a)}
	\end{minipage}
	\begin{minipage}[b]{0.48\linewidth}
		\includegraphics[width=1\linewidth]{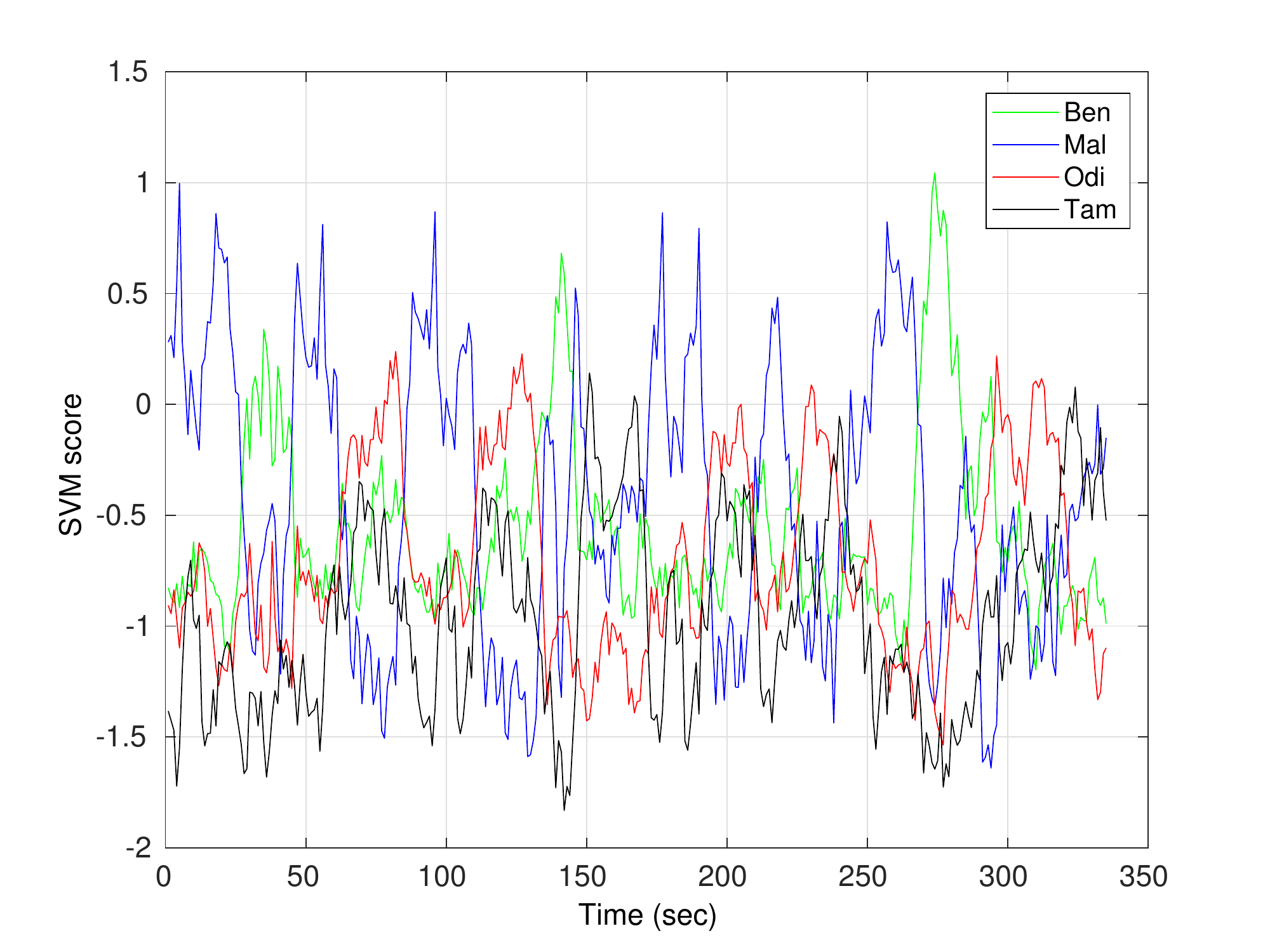}
		\centerline{(b)}		
	\end{minipage}
	\begin{minipage}[b]{0.48\linewidth}
		\includegraphics[width=1\linewidth]{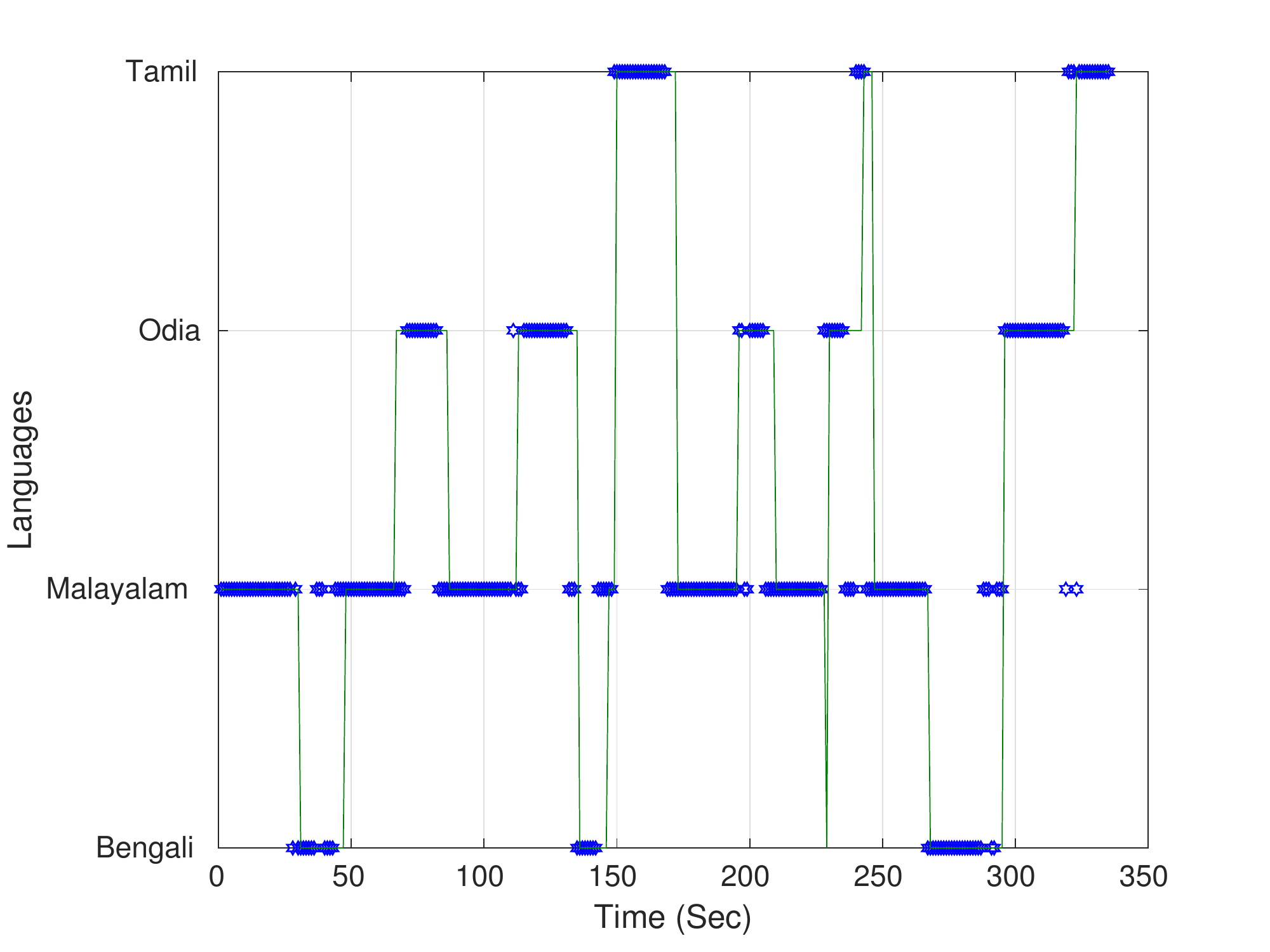}
		\centerline{(c)}
	\end{minipage}
	\begin{minipage}[b]{0.48\linewidth}
		\includegraphics[width=1\linewidth]{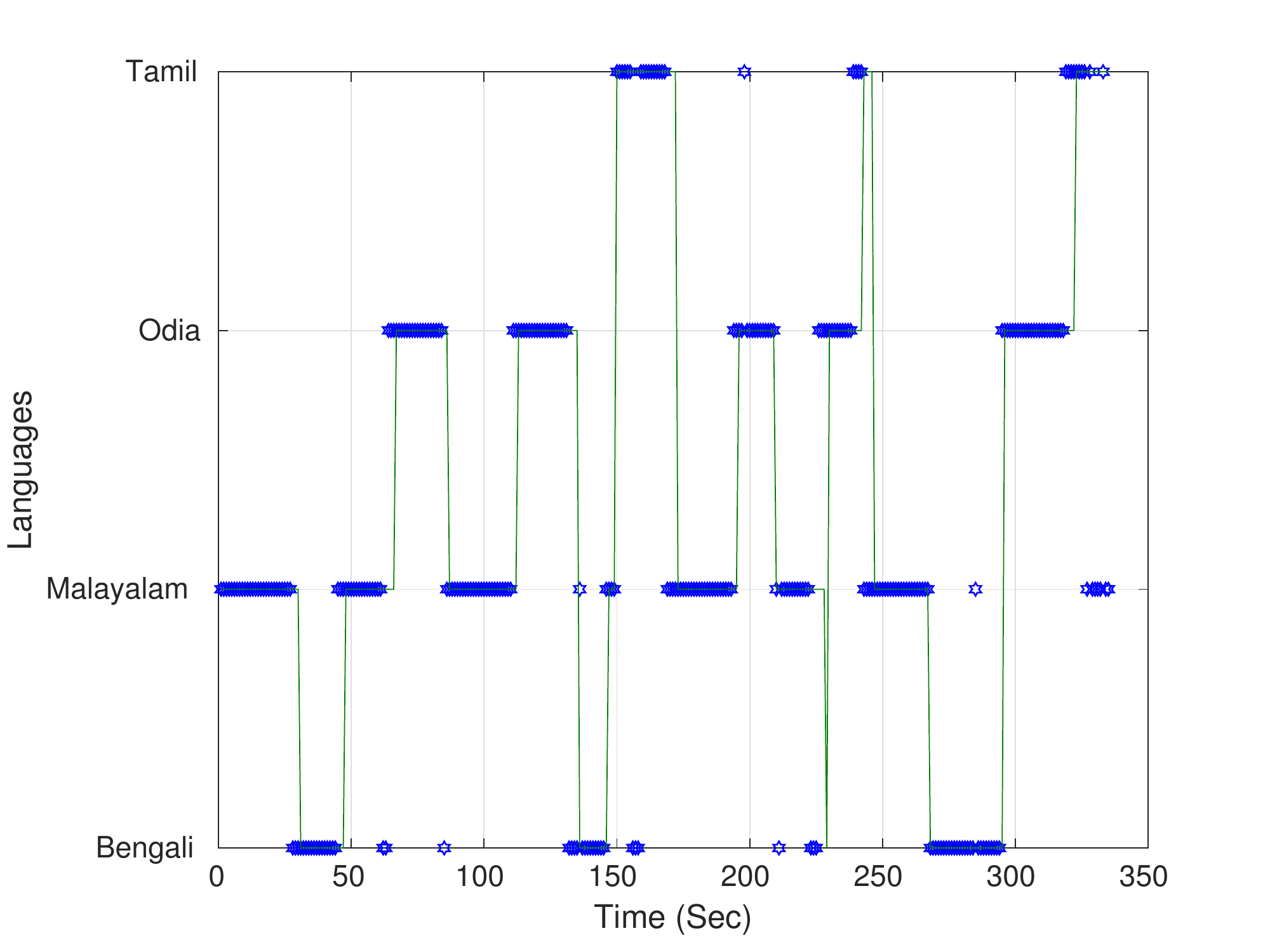}
		\centerline{(d)}		
	\end{minipage}
\caption{language based speech segmentation (a) SVM score for each language with scheme-1 (b) SVM score for each language with scheme-2  }\label{fig:four_lang}
\end{figure}

 
From the result, we can see that the segmentation performance is poor with ten language training model because the chances of misclassification among the language is more. Hence, we have also evaluated the language segmentation performance by taken four specific language training model. We have taken the above mentioned sub-group-I languages and build the training model by applying scheme-1 and scheme-2. We have built the synthetic concatenated conversion with this four languages by taking 3 segments per language and the segment duration is in between 6-30 seconds with 2-5 seconds of sliding window with shift of 1 second. We have found that with four language model, scheme-1 has performed better than scheme-2. The results have been shown in Table \ref{Table.four}. In the Fig.\ref{fig:four_lang} 
\begin{table}[!h]
\centering
\resizebox{0.3\textwidth}{!}{%
\begin{tabular}{|l|l|l|l|}
\hline
\begin{tabular}[c]{@{}l@{}}Sl. \\ No\end{tabular} & \multicolumn{1}{c|}{\begin{tabular}[c]{@{}c@{}}Sliding \\ Window Size (sec)\end{tabular}} & \multicolumn{1}{c|}{\begin{tabular}[c]{@{}c@{}}Scheme \\ 1 (\%)\end{tabular}} & \multicolumn{1}{c|}{\begin{tabular}[c]{@{}c@{}}Scheme \\ 2 (\%)\end{tabular}} \\ \hline
1.                                                & 2                                                                                         & 64.84                                                                           & 67.63                                                                           \\ \hline
2.                                                & 3                                                                                         & 72.44                                                                           & 69.36                                                                           \\ \hline
3.                                                & 4                                                                                         & 75.31                                                                           & 71.81                                                                           \\ \hline
4.                                                & 5                                                                                         & 78.13                                                                           & 73.14                                                                           \\ \hline
\end{tabular}}
\caption{Comparison of language segmentation accuracy between two approaches}\label{Table.four}
\end{table} 
\section{Conclusion}
In this paper, we have explored the language identification and segmentation work in Indian regional language context. We have explored two schemes to understand the performance of individual language with respect to different feature variation. We have used difference matrix and n-gram based features with SVD based feature embedding. We have observed depending upon the test duration the performance varies between the schemes. In this study, we have also highlighted individual language based inference from the result. 
We have also seen that segmentation accuracy of a group of Indo-Aryan languages are better than group of Dravidian languages that means similarities between the Dravidian languages are more than the Indo-Aryan languages. Also, as expected segmentation accuracy is more with same language training model rather than training with the complete language model. In future we may extend this work with deep neural network approaches. 
\bibliographystyle{IEEEbib}

\end{document}

%% file: train_diagram.tex
\begin{figure}[!h]
\centering
 \resizebox{0.45\textwidth}{!}{%
\begin{tikzpicture}
  \node[block] (a) {AIR Language Database (10 Indian languages)};
  \node[block] (b) at ([yshift=-2cm]$(a)$) {Preprocessing (Remove music and silences)};
  \draw[line] (a)-- (b);
  \node[block1,right=3cm of b] (c) {Feature Extraction \& CMN};
 
   \path [line] (b) -- node [text width=2.5cm,midway,above,align=center] {3 mins of each utterance} (c);
    \draw[line] (b)-- (c);
   \node[block1,right=of c] (d) {UBM};
   \draw[line] (c)--(d);
     \node[block1] (e) at ([yshift=-3cm]$(b)$) {Feature Extraction \& CMN};
     \draw[line] (b)--(e);
     \path [line] (b) -- node [text width=2.5cm,left,align=center] {1 mins of each utterance} (e);
     \node[block1,right=3.5cm of e] (f) {MAP adaptation};
     \draw[line] (e)--(f);
     \draw[line] (c)--(f);
     \node[block1,right=2cm of f] (g) at ([yshift=1.2cm]$(f)$) {Decoding Sequence};
     \node[block1,right=2cm of f] (h) at ([yshift=-1.2cm]$(f)$) {Difference Matrix};
     \draw[line] (f)--(g);
     \draw[line] (f)--(h);
      \node[block2,right=1cm of g] (i) {Ngram Transition Matrix \& SVD};
      \node[block2,right=1cm of h] (j) {Feature Embedding (SVD)};
     \draw[line] (g)--(i);
     \draw[line] (h)--(j);
     \node[block1,right=1cm of i] (k) {SVM Modelling};
     \draw[line] (i)--(k);
     \node[block1,right=1cm of j] (l) {SVM Modelling};
     \draw[line] (j)--(l);
     \node[block1,right=1cm of k] (m) {Trained Model 1};
     \draw[line] (k)--(m);
     \node[block1,right=1cm of l] (n) {Trained Model 2};
     \draw[line] (l)--(n);
     \node[draw,dotted,fit=(g) (i) (k) (m)] (s) {};
     \node[draw,dotted,fit=(h) (j) (l) (n)] (t) {};
     \node[line,right=1cm of e] at ([yshift=-0.85cm]$(s)$) (u) {Scheme 1};
     \node[line,right=1cm of e] at ([yshift=-0.85cm]$(t)$) (u) {Scheme 2};
\end{tikzpicture}}
\caption{Language Identification/Segmentation Schemes.}
\label{fig:train_dia}
\end{figure}

%% file: test_dia.tex
\begin{figure}[!h]
\centering
 \resizebox{0.45\textwidth}{!}{%
\begin{tikzpicture}
  \node[block1] (a) {Test Language Utterance};
  \node[block1,right=1cm of a] (b) {Feature Extraction \& CMN};
  \node[block1,right=1cm of b] (c) {MAP Adaptation};
  \node[block1] (d) at ([yshift=2cm]$(c)$) {UBM};
  \node[block1,right=2cm of c] (e) at ([yshift=1.2cm]$(c)$) {Symbol Sequence and N-gram Model};
   \node[block1,right=2cm of c] (f) at ([yshift=-1.2cm]$(c)$) {Difference Matrix};
   \node[block2,right=1cm of e] (g) {Feature Embedding \& SVD};
   \node[block2,right=1cm of f] (h) {Feature Embedding \& SVD};
   \node[block,right=1cm of g] (i) {Decision};
   \node[block,right=1cm of h] (j) {Decision};
   \node[block1] (k) at ([yshift=2cm]$(i)$){Trained Model 1};
  \node[block1] (l) at ([yshift=-2cm]$(j)$){Trained Model 2};
  \draw[line] (a)--(b);
  \draw[line] (b)--(c);
  \draw[line] (d)--(c);
  \draw[line] (c)--(e);
  \draw[line] (c)--(f);
  \draw[line] (e)--(g);
  \draw[line] (f)--(h);
  \draw[line] (g)--(i);
  \draw[line] (h)--(j);
  \draw[line] (k)--(i);
  \draw[line] (l)--(j);
\end{tikzpicture}}
\caption{Testing Phase}
\label{fig:test_dia}
\end{figure}

%% file: lang_dia.tex
\begin{figure}[!h]
\centering
 \resizebox{0.45\textwidth}{!}{%
\begin{tikzpicture}
\node[inner sep=0pt] (a) at (0,0)
    {\includegraphics[width=.25\textwidth]{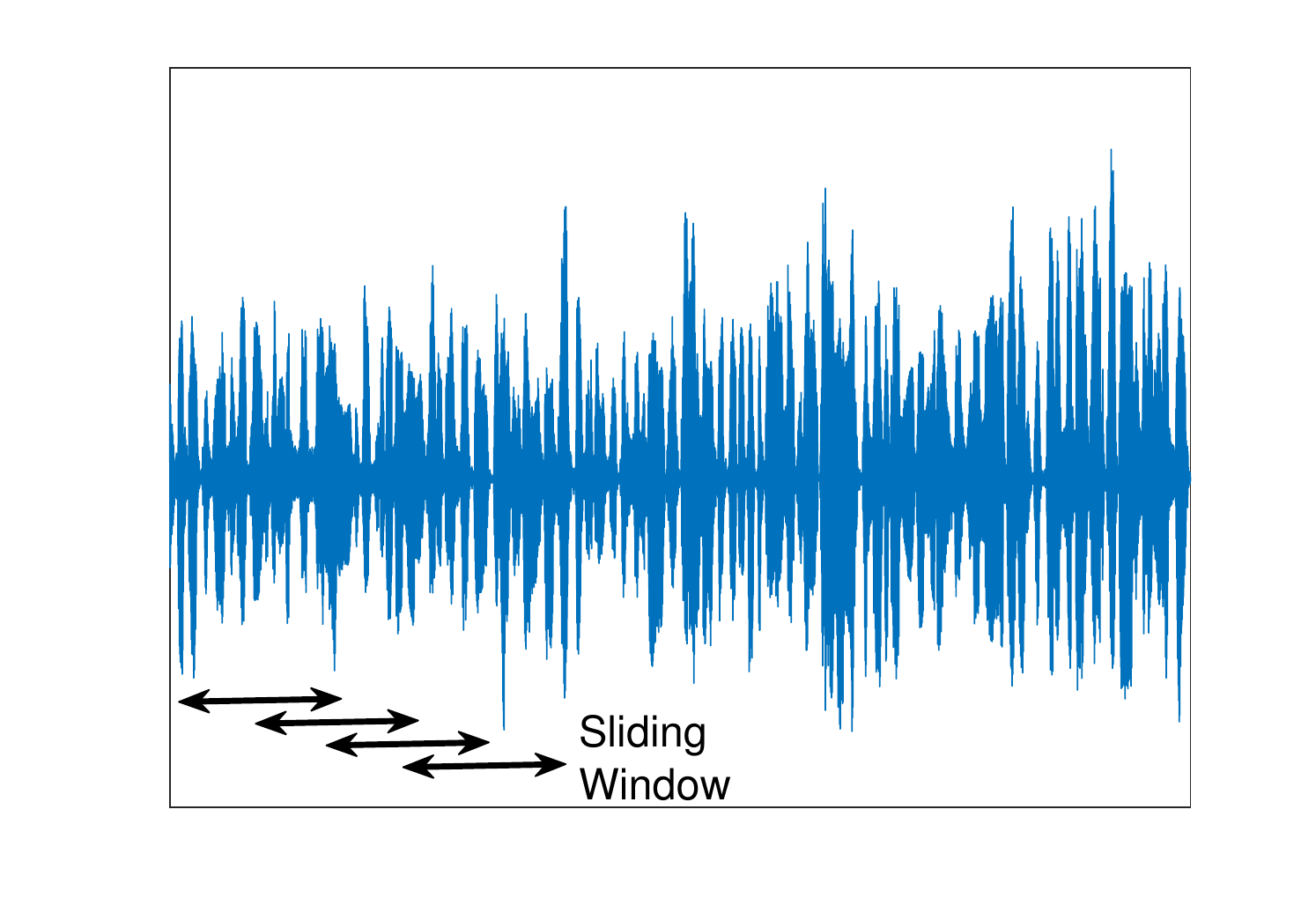}};
     \node[block1] (b) at ([xshift=3cm,yshift=-2cm]$(a)$) {Scheme-1};
     \draw[line] (a)|- (b);
     \node[block1] (c) at ([yshift=-4cm]$(b)$) {Scheme-2};
     \draw[line] (a)|- (c);
     \node[inner sep=0pt,right=of b] (d)
    {\includegraphics[width=.25\textwidth]{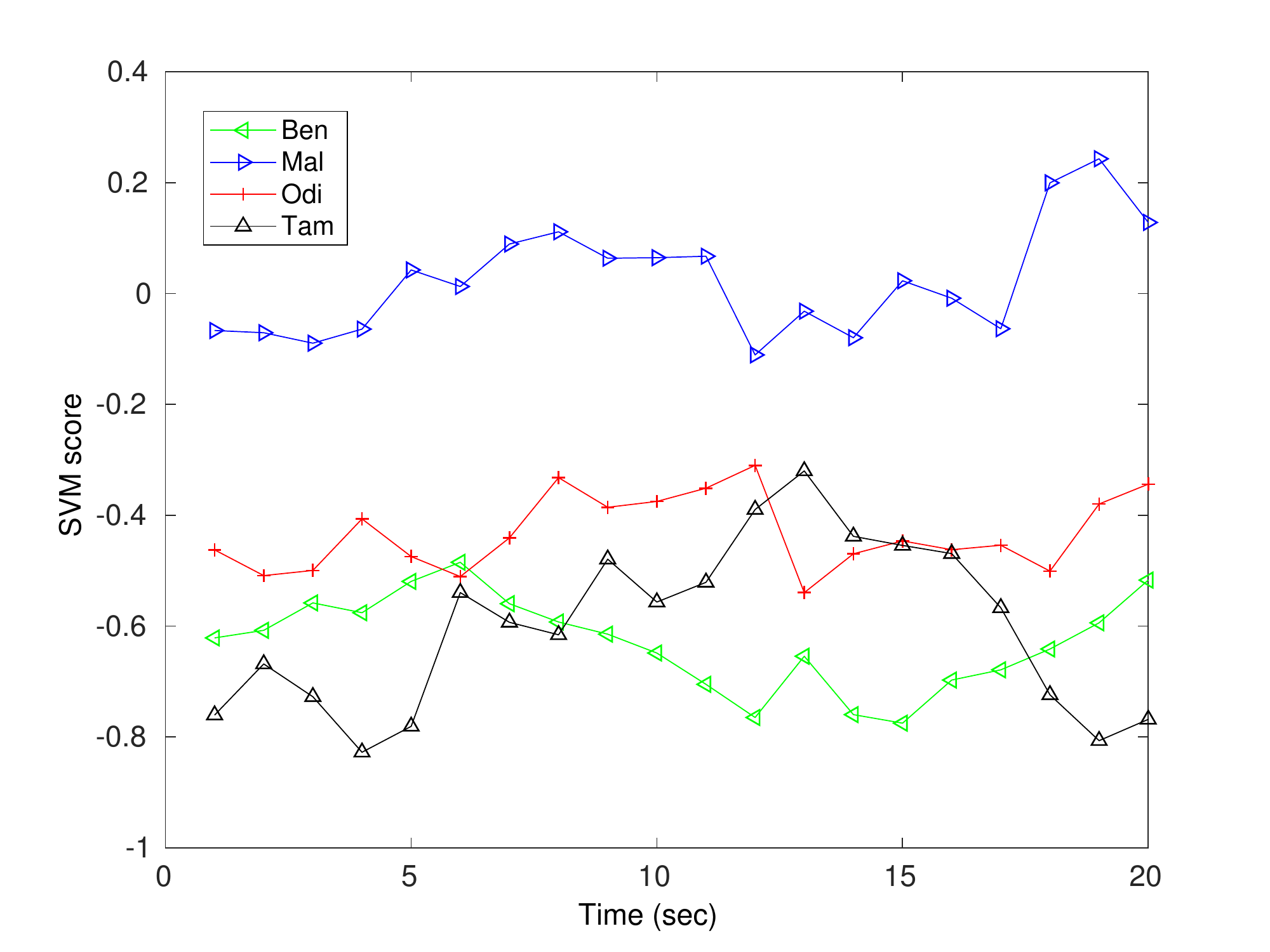}};     
    \node[inner sep=0pt,right=of c] (e)
    {\includegraphics[width=.25\textwidth]{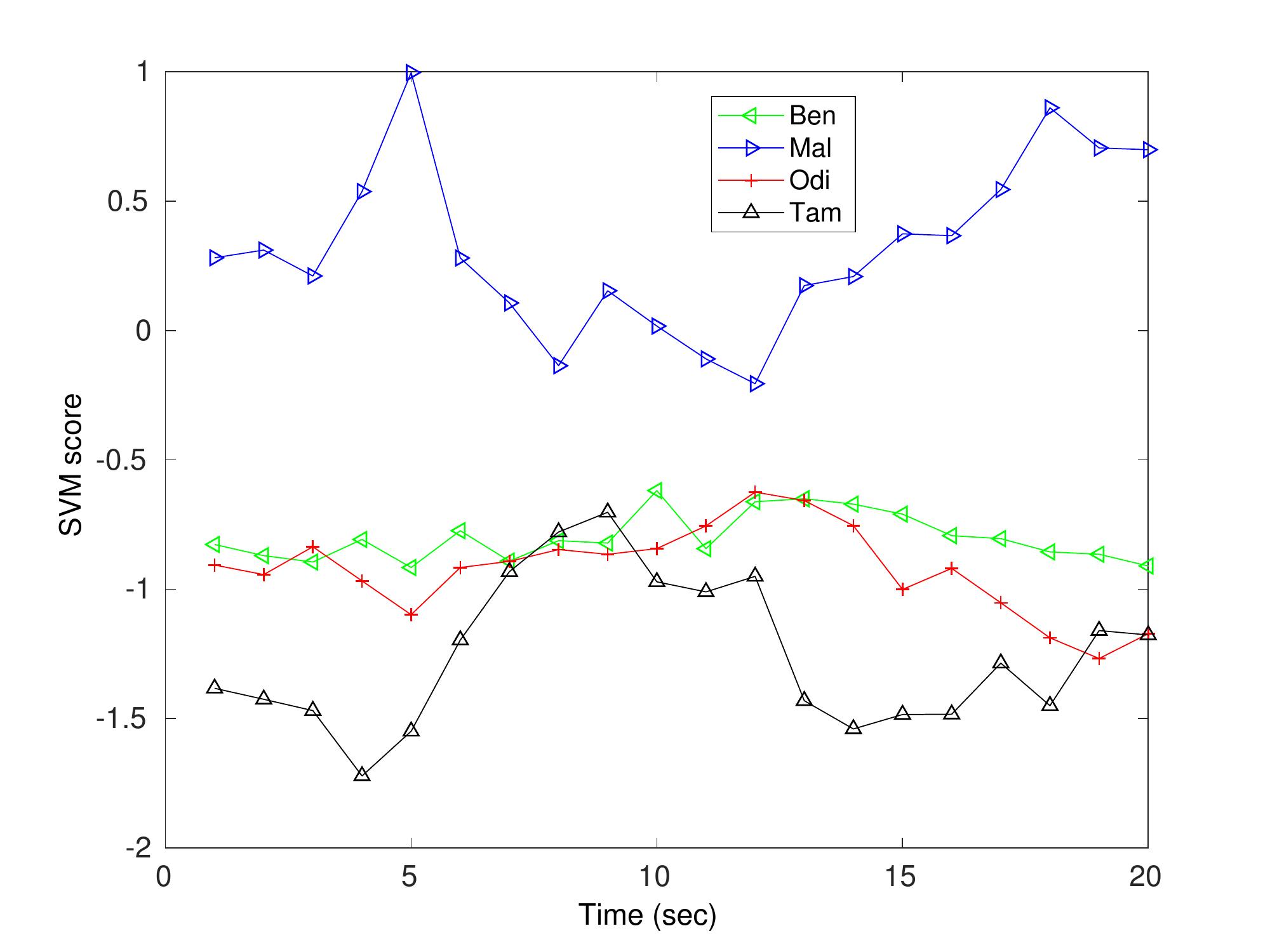}};
    \draw[line] (b)|- (d);
    \draw[line] (c)|- (e);
\end{tikzpicture}}
\caption{Language segmentation procedure}
\label{fig:lang_dia}
\end{figure}